\documentclass[a4paper,12pt,reqno,superscriptaddress,nofootinbib]{revtex4}
\usepackage[centertags]{amsmath}
\usepackage{amsfonts}
\usepackage{amssymb}
\usepackage{amsthm}
\usepackage{newlfont}
\usepackage{stmaryrd}
\usepackage{mathrsfs}
\usepackage{mathtools}
\usepackage{euscript}
\usepackage{graphicx}
\usepackage{enumerate}
\usepackage{todonotes}
\usepackage[normalem]{ulem} % for strikeout text with \sout
\usepackage{comment}
\usepackage{color}
\usepackage{floatrow}

\DeclareMathOperator{\Tr}{Tr}

\newcommand{\id}{\textrm{d}}

   \let\ep=\epsilon

\newcommand{\bbR}{{\mathbb R}}

\usepackage{tikz}
\usepackage{pgf}
\usetikzlibrary{positioning,fit,calc}
\usetikzlibrary{arrows,automata}
\usepackage{wrapfig}
\usepackage{subfigure}
\usepackage{amscd}
\usepackage{hyperref}

\theoremstyle{plain}
  \newtheorem{theorem}{Theorem}[section]

\theoremstyle{definition}

\theoremstyle{remark}
  \newtheorem{remark}[theorem]{Remark}

  \newtheorem{example}[theorem]{Example}

   \let\ep=\epsilon

\newcommand{\bbC}{{\mathbb C}}

\newcommand{\opunit}{\text{1}\kern-0.22em\text{l}}

\DeclareMathAlphabet{\mathpzc}{OT1}{pzc}{m}{it}

\let\oldsqrt\sqrt
\def\sqrt{\mathpalette\DHLhksqrt}
\def\DHLhksqrt#1#2{%
\setbox0=\hbox{$#1\oldsqrt{#2\,}$}\dimen0=\ht0
\advance\dimen0-0.2\ht0
\setbox2=\hbox{\vrule height\ht0 depth -\dimen0}%
{\box0\lower0.4pt\box2}}

\begin{document}

\title{Response theory: a trajectory-based approach}

\author{Christian Maes\\{\it Instituut voor Theoretische Fysica, KU Leuven}}

\begin{abstract}
	We collect recent results on deriving useful response relations also for nonequilibrium systems. The approach is based on dynamical ensembles, determined by an action on trajectory space.  (Anti)Symmetry under time-reversal separates two complementary contributions in the response, one entropic the other frenetic.  Under time-reversal invariance of the unperturbed reference process, only the entropic term is present in the response, to give the standard fluctuation--dissipation relations in equilibrium.   For nonequilibrium reference ensembles, the frenetic term contributes essentially and is responsible for new phenomena.  We discuss modifications in the Sutherland-Einstein relation, the occurence of negative differential mobilities and the saturation of response. We also indicate how the Einstein relation between noise and friction gets violated for probes coupled to a nonequilibrium environment.  We end with some discussion on the situation for quantum phenomena, but the bulk of the text concerns classical mesoscopic (open) systems.\\
	The choice of many simple examples is trying to make the notes pedagogical, to introduce an important area of research in nonequilibrium statistical mechanics.
\end{abstract}
\maketitle
%\indent\hspace{4cm} {\tiny Stri Parva  --- The Mahabharata, Translated by Kisari Mohan Ganguli (1889), Ch. 2, p6.}
\vspace{0.6cm}
\newpage
\tableofcontents

\newpage
\section{Introduction}
To know a system operationally, is to be able to predict its response to a stimulus. Conversely, we learn about a system by observing its response. In many ways and in all sciences, that is the very ground for doing experiments where we interfere with the system's condition.   When, in psychology, subjects are tested for their reaction to external stimuli, conclusions are being drawn about susceptibility or vulnerability.  In other domains from sociology to climate science, we speak of the impact of events or measures, and/or of resilience of the system of interest; see e.g. \cite{cvca}.  For biological processes, adaptation (i.e., {\it proper} response) to changes in the environment  is a matter of survival. How robust are foodwebs or other (economic) networks over which supply and demand move?   On micro-scales, mechanotransduction makes  cells respond biochemically to mechanical stimuli.  All of these areas are of immense interest and even importance today.\\
 In physics and since a long time, response has been associated with transport phenomena. The transport of particles, energy, volume or momentum is a central subject in all of physics.  Pushing, driving, stimulating or exciting a system in one or the other way, leads to displacements in physical quantities. The amount and nature of any displacement and how it depends on the original condition is the subject of response theory.  Transport coefficients such as conductivities and mobilities, viscosities and elasticity moduli, have therefore been studied often in the context of response theory.\\
   Over time however, a more general framework has emerged, to begin with linear response theory around equilibrium.  It is the context of so called fluctuation--dissipation relations. The terminology hints at the nature of the result, at least for equilibrium systems: response got connected with fluctuating quantities, in some cases expressing dissipation or diffusion of quantities like energy, position or velocity.  As a consequence, response theory also played a role in summarizing or establishing irreversible behavior on macroscopic scales starting from reversible microscopic laws; see e.g. \cite{fourier}.\\
   Response theory for systems out-of-equilibrium is of more recent times.  One major problem, even for the more restricted class of nonequilibrium processes considered here, is that the response is no longer describable in terms of thermodynamic variables like energies or entropy.  Kinetics enters and the steady condition is not characterized simply in terms of a few macroscopic quantities.  Typically we do not know the stationary distribution, and yet we wish to formulate response in terms of observable quantities.  This is the main attempt of the paper, to explain an approach to response which is trajectory-based, meaning to formulate ensembles on the space of allowed trajectories.  The action or Lagrangian contains both thermodynamic and kinetic information about the process, and that gets reflected in response relations.   The trajectory-based approach of the present paper, on micrometer scales, is compatible with the recent great progress in monitoring and manipulating mesoscopic trajectories of tagged particles.  We have in mind fluorescence and fast-camera tracking, combined with optical manipulations and shaping of potentials and driving, e.g. via optical tweezers (1986) \cite{ash}.  Such experimental tools enable to collect also kinetic (and not only thermodynamic) information, which appears an unavoidable prerequisite for understanding nonequilibrium behavior.\\ 
From the conceptual point of view, we must prepare the scene and introduce structure in (nonequilibrium) response. From what will follow below, the most important players to correlate with are excesses in entropy flux and frenesy.  The last concept is relatively new, and requires examples and illustrations to understand its operational meaning.  In  particular, response measurements will give information about changes in dynamical activity and escape rates, which constitute the meaning of frenesy.  We refer to recent monographs on frenesy for an update, \cite{fren,springer}.     In all, we seek expressions of response that are informative or operationally useful.  Response theory indeed hopes to relate the stimulus with observable effects in the unperturbed system. The ambition is thus bigger than providing a Taylor expansion or some formal perturbation series in the amplitude of the stimulus.  Understanding response means to identify mechanisms and specify observables that are relevant even independent of the detailed model, stimulating intuition and enabling to reconstruct the response in terms of some more elementary considerations.  % But also on atmospheric, stellar and even cosmological scales, response may be expressed via fluctuation relations whenever those fluctuations matter and become observables (thinking for example of the cosmic microwave background pictured by  Planck and WMAP missions).
\\

\begin{figure}[!h]
	\centering
	\includegraphics[width=0.55\textwidth]{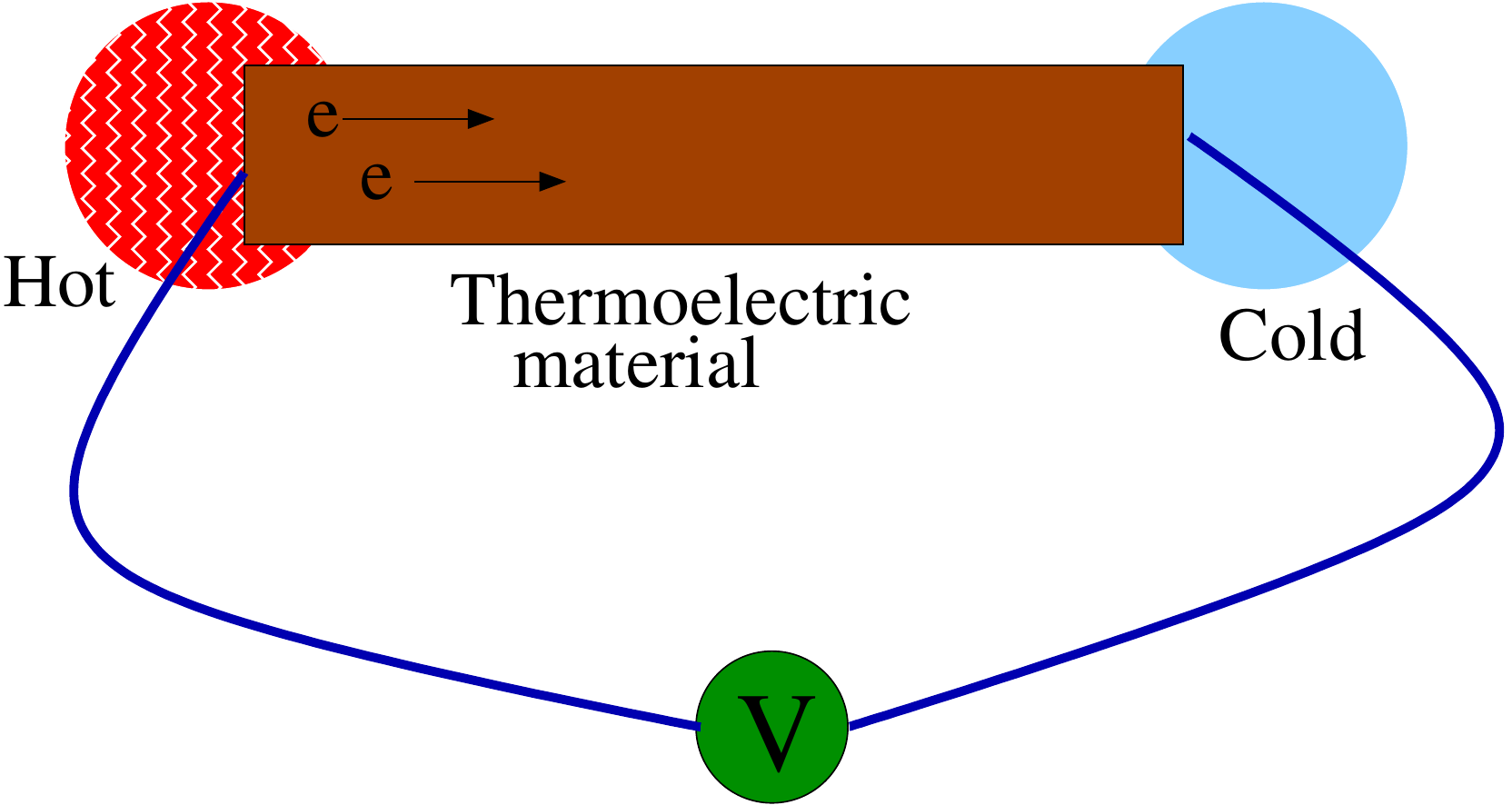}
	\caption{Seebeck-Peltier effect.  Electric and thermal currents are the response to small voltage and temperature differences.  Their interference is described by symmetric Onsager transport coefficients.}
	\label{fig:the}
\end{figure}

Response relations have been formulated since a very long time, and their contents never failed to impress.   An early example has the typical setup drawn in Fig.~\ref{fig:the}.  It concerns the second Thomson relation (1854) between the Seebeck and the Peltier coefficients.  Their equality was understood to be a manifestation of time-reversal invariance in the 1931-work \cite{O} of Lars Onsager.  Such Onsager reciprocity relations as indeed found in thermoelectric phenomena are useful to decrease the number of unknown linear response coefficients.  They can also be read off from the Green--Kubo relations that were derived hundred years after the paper by Kelvin \cite{kel}. The general idea is that in linear response around equilibrium, the average of a current $\langle J_i\rangle^F$ of type $i$ (e.g. an electric current) is proportional to its correlation with the excess entropy flux $S$,
\begin{equation}\label{greku}
\langle J_i \rangle^F = \frac 1{2}\langle S\,J_i\rangle = \frac 1{2}\sum_k \langle J_i\,J_k\rangle\,F_k, \quad S = \sum_k J_k\,F_k
\end{equation}
where $F_k$ is the thermodynamic force of type $k$ (e.g. giving the difference in temperature at opposite ends of the system). The linear response coefficients $\langle J_i\,J_k\rangle$ with averages in the equilibrium ensemble are clearly symmetric under exchanging $i\leftrightarrow k$ (e.g. allowing to identify the Seebeck with the Peltier coefficient divided by temperature). 

\begin{figure}[!h]
	\centering
	\includegraphics[width=0.45\textwidth]{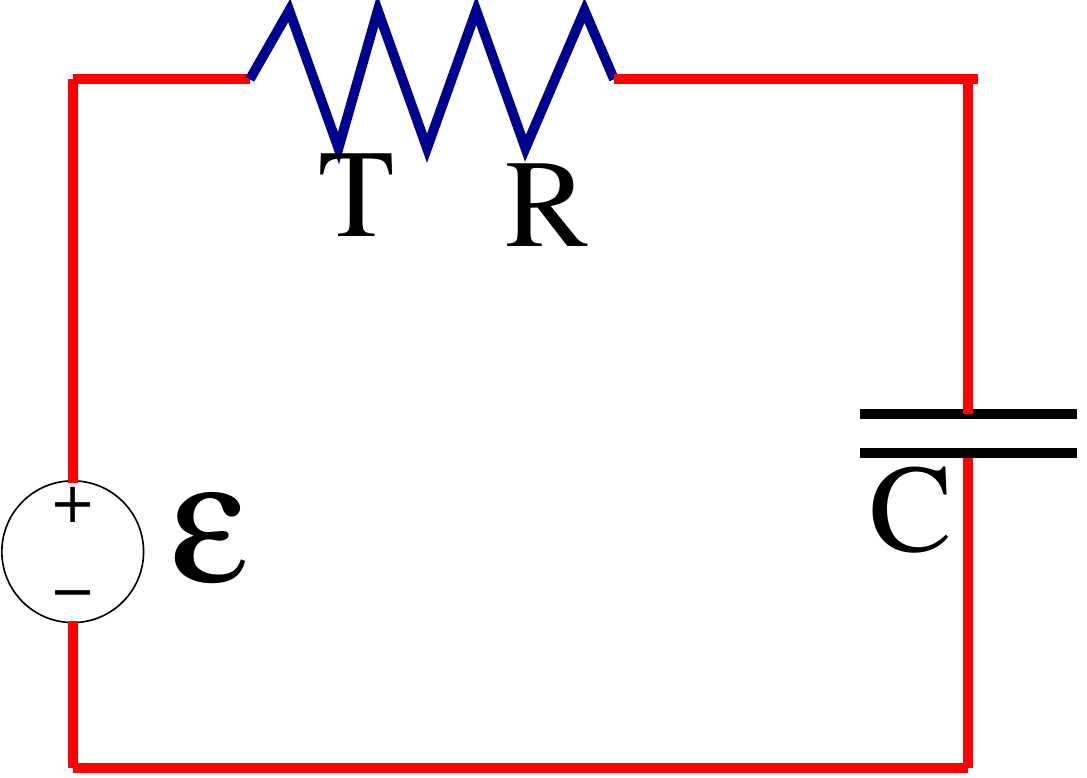}
	\caption{RC-cicuit with resistor at temperature $T$, the electrical (linear) circuit equivalent of the basic Langevin equation.  The thermal noise at the resistor produces a fluctuating potential, following the Johnson--Nyquist effect.}
	\label{fig:cir}
\end{figure}

\noindent(We ignore for the moment the issue of parity and generalized Casimir-Onsager reciprocity.)  The intervention of the entropy flux, defined from a balance equation, was in essence the start of much of irreversible thermodynamics \cite{dGM}.\\
Another line of response theory started with the PhD work of Pierre Curie  (1896) on the magnetic susceptibility of paramagnets. There, we do not deal with transport or with currents but we look at the response of magnetization. Curie derived that at high temperature the equilibrium magnetization $m_h$ responds to a small external magnetic field $h$ with susceptibility 
$\chi$, for which
\[
m_h - m_0 = h \;\chi, 
\qquad \chi \sim \frac 1{T} 
 \]
I.e., the magnetic susceptibility falls off with the inverse of the absolute temperature $T$ (law of Curie).
% \begin{figure}[!h]
%	\centering
%	\includegraphics[width=0.45\textwidth]{magnetization.eps}
%	\caption{Cartoon of nonequilibrium response}
%	\label{fig:mag}
%\end{figure}
 The structure of such relations has been clarified by the Gibbs formalism, where free energies govern responses via their derivatives.  E.g., heat capacities are thereby related to variances in energy or enthalpy.  Mixed derivatives give rise to an analogue of the Onsager reciprocity for linear transport coefficients known as Betti-Maxwell reciprocity (in equilibrium elasticity theory).\\
  Perhaps the best-known response formula however is the Sutherland-Einstein relation (1904--05), \cite{suth,ein}.  There, the mobility is proportional to the diffusion constant.   It is a functional cornerstone of much of colloidal physics.  We will see various elementary examples in Section \ref{exam}.  All of the above are called fluctuation--dissipation relations of the first type.\\
    A further line of relations, following from response theory and called  fluctuation--dissipation relations of the second type, has been opened by the Johnson-Nyquist formula. It gives an expression for the noise arising from the thermal
agitations of the electrons in a resistor.  As a consequence,
a random voltage emerges which can be measured at the ends of the
resistor (Johnson effect, 1926).  Mathematically, that voltage can be described as the
random voltage source $U^f_t$ given in the Nyquist formula (1928),
\begin{equation}\label{nyq}
U^f_t = \sqrt{2k_BT\, R} \;\xi_t
\end{equation}
with $R$ the resistance and $\xi_t$ a standard white noise.  The amplitude is of
course very small by the presence of Boltzmann's constant $k_B$, at least when compared to macroscopic
voltage values. Representing each resistor as an ideal resistor in series with the source \eqref{nyq}, we can study fluctuations in an arbitrary electrical circuit. As an example, consider a resistance $R$ in series with a capacity $C$ and with a steady voltage source $\cal E$; see Fig.~\ref{fig:cir}.

\noindent Write $U_t$ for the variable potential difference over the capacitor. Kirchhoff's second law
reads
\begin{equation}\label{rclang}
RC\,\dot U_t = \cal E - U_t + U_t^f
\end{equation}
By inserting the white noise $\xi_t$ following~\eqref{nyq}, we obtain the Langevin equation
\begin{equation}\label{rclang1}
\dot U_t = \frac{\cal E - U_t}{R C} + \sqrt{\frac{2k_BT}{R C^2}}\, \xi_t
\end{equation}
With the battery removed,
$\cal E = 0$, the dynamics is reversible for energy function $H(U) = C U^2/2$. In particular,
$\lim_{t\uparrow\infty} \langle U_t^2 \rangle = k_BT/C$, in accordance with the
equipartition theorem.  We can however also see from \eqref{rclang1} how the potential changes when the battery is turned on or when $\cal E$ changes in time.  That is again the subject of response theory and the answer obviously depends on and should make use of the choice \eqref{nyq}; we come back to that example at the end of Example \ref{lav}

From the above (more historical) examples we already become aware of a possible connection between response and dissipation as expressed in fluctuation relations.
That will be systematized in the following sections.  In that respect it is useful to keep distinctions clear and to separate various questions.  Terminology is not always helpful here, as such terms as fluctuation--dissipation relations, Einstein relation, response relation etc. are used in multiple meanings throughout the literature.

\section{General question and ambitions}\label{x}

Response will be collected in a time-interval $[0,t]$.  At negative times $s
\leq 0$ (all the way to time zero) the system of interest has been prepared in a reference condition.  That can be many things, from a thermal equilibrium condition to a specific transient regime or, most often in this paper, a steady nonequilibrium reference.  The idea is that at time zero, the system (in whatever prepared or reference condition) opens to a time-dependent stimulus.  That stimulus will be treated as a perturbation and hence we speak of linear versus nonlinear response depending on the sought consequence of the (small) stimulus.  Both the stimulus (or perturbation) and the observed quantity are allowed to be time-extensive over $[0,t]$; see Fig.~\ref{fig:res}.\\ 
The goal of response theory is to describe and predict in a systematic and physical way the statistical response, preferably from observations that could be made in the initial (reference) condition.  The word ``statistical'' refers to the fact that we deal with a reduced description, physically compatible with the microscopic laws but on a level where the hidden degrees of freedom have been integrated out (after some infinite volume limit, in weak coupling etc) and provide ``enough'' noise for dissipative behavior.  In that respect it is not necessarily the task of response theory to \emph{demonstrate} dissipative behavior; rather, its validity will depend on it.\\  

As is clear from scanning the vast literature on the subject, there are many different versions of response theory.  Apart from standard treatments in text books such as \cite{ku,chan,ba,ma,zwa}, they include  the papers \cite{res1,res2,res3,res4,res5,res6,res7,res8,res9,res10,res11,res12,res13,srlin,viv,sar2} to which we refer for other approaches and results.  The originality of our approach is to start from dynamical ensembles on path-space. The action governing the weight of a trajectory will get a physical significance in its decomposition in a time-antisymmetric source (entropy flux) and a time-symmetric contibution (frenesy) which both change due to the perturbation. The merit of response theory is indeed not its formal appearance  -- in the end we are all doing Taylor expansion assuming (and sometimes proving) convergence of certain integrals. In particular, for nonequilibrium purposes,  we emphasize the importance of the frenetic contribution in response; for different details and discussions, we refer to \cite{fdr,njp,urna,alb,pccp,fdr2,fdr1}.\\

We end those verbosities by winding three final remarks around the main subject:
\begin{remark}\label{vk}
- \underline{The objection by  Nico van Kampen (1971)}  against linear response theory and the derivation of (Green--)Kubo relations has been widely discussed.  The original concerns were formulated in \cite{vk}.  Multiple reactions and answers have been given. To summarize the situation, van Kampen criticised the microscopic approach via the Liouville equation (which one still often encounters in text books and reviews).  Linearizing the microscopic theory is no justification of linear macroscopic equations (with currents proportional to forces).  Moreover, microscopic dynamics can be very nonlinear in the sense of possessing strong dependence on initial conditions.  Linear response on that micro-level would only hold for absurdly short times.\\
These objections are of course fully justified, but linear response need not proceed so naively as criticized by van Kampen.  In a way, and in no contradiction with van Kampen's objection, linear response can only be expected to work well on scales of descriptions where ``noise has been effective'' to make the reduced description sufficiently chaotic.  Paradoxically, instabilities typically help to assure sufficient statistical mixing; see also \cite{cha}.  In what we will discuss, the system is open and assumed to be described by a probability law on trajectories with an action which is sufficiently local in spacetime.  A simple realization are Markov processes.  The physics that proceeds that description is one of weak coupling with an infinite bath of components which evolve on a much faster time scale.   The correct order of linear response is indeed to first take the thermodynamic limit and to focus on a reduced description which is sufficiently spacetime-mixing.  Then, only afterwards, the limit of linear response can be taken.  Linear reponse formul{\ae} will therefore not {\it prove} diffusive or dissipative behavior on meso- to macroscopic scales, but instead {\it depend} on it for their full justification. 
\end{remark}
\begin{remark}- \underline{The issue of causality and relaxation} amounts to the question whether we should \emph{impose} or rather \emph{derive} the fact that the response happens after the stimulus.  It would seem natural that no extra condition of causality is needed; the dynamics with its perturbation should take care of that.  That is also the option we are taking. Nevertheless, the fact that {\it it takes time} for a perturbation at some fixed moment to relax away so that the system may return to its original condition, is deep and interesting even in classical physics. Clearly, estimating relaxation times is not purely a question of thermodynamics. That convergence is fast enough requires absence of jamming and localization. Response theory indeed uses time-correlation functions and their (sufficient) decay is an assumption or a result whose justification falls outside response theory all together. Stability of (non)equilibria \cite{gp} is a subject which is clearly related to response theory but the latter often pre-supposes the first.
\end{remark}\begin{remark}
- \underline{Numerical work} and in particular equilibrium molecular dynamics has been succesfully used  to compute transport coefficients from the Green-Kubo formulae. For nonequilibrium response relations, various new algorithms, in particular using thermostated dynamics, have been employed.  Numerical methods and their physical motivation fall out of the scope of the present discussion but we refer to the book \cite{eva} for more material and references. For nonequilibrium response, the search for efficient numerical algorithms to evaluate the FDR such as the so-called zero-field (or field--free) algorithms played an important role; see \cite{chat,ric,res12,cor}. 

\end{remark}

\subsection{Plan of the paper}
After presenting a number of well-known and more elementary examples, we introduce the main formal tool in Section \ref{dynens}.  Dynamical ensembles are presented with their action and decomposition in time-symmetric and time-antisymmetric excesses.  There will be plenty of examples to illustrate their nature for various types of Markov processes satisfying local detailed balance.  As such however, dynamical ensembles may stand on their own and do not essentially depend on specifying the underlying dynamical equations.  The response theory in Section \ref{res} depends mathematically solely on the action.  Its decomposition becomes meaningful by giving rise to two major contributions to the response (entropic and frenetic). The discussion on response around nonequilibria makes the main part of the paper, but we also discuss a unifying view on response around equilibria. Apart from presenting various cases of response, we also explain the relation with local detailed balance and the different kinds of fluctuation-dissipation relations that exist.  We often concentrate there on the Sutherland--Einstein relation and its possible violation.  We discuss some experimental challenges and higlight the Harada-Sasa equality.  In Section \ref{nl} we give examples of response relations for active particles, where local detailed balance does not hold.  We end with the quantum case in Section \ref{qca}, both as a reminder of what is true and to open the question for trajectory-based versions of quantum nonequilibrium response.\\
Having stated that, there are naturally also many things which are not being discussed explicitly in the present review.  There is for example no discussion on nonequilibrium additions to viscosities and elastic moduli, hence not touching the subject of \emph{odd} viscocity and elasticity 
\cite{av,chic,calt,calt1}.  We also spend very little time with aspects of heat conductivity and with the question of anomalous transport (in low dimensions), see e.g. 
\cite{liv} and references therein.  In particular, we do not address the question of integrability of (even) Kubo expressions for the linear transport, when there are more conserved quantities, when (almost) integrability obtains or in low dimensions.  All of those are important topics of current research but here we have chosen to highlight only the most elementary structures in a pedagogical exposition for making the bridge to nonequilibrium response theory.

\subsection{Elementary examples}\label{exam}

\begin{example} [Langevin dynamics]\label{lav}
Consider a small particle of mass $m$ in a thermal environment at temperature $T$.  At time $s=0$ an external force field $F_s$ is turned on.  From then the dynamics is modeled with the perturbed Langevin evolution (in one dimension) with position $q_s$ following $\dot{q}_s=v_s$ and velocity $v_s$ changing with 
\begin{equation}\label{lvin}
\dot v_s = -\gamma\,v_s + \frac 1{m}\,F_s + \sqrt{2k_BT \gamma/m}\,\xi_s,\qquad s> 0
\end{equation}
where (here and later) $(\xi_s)_s$ is a standard white noise process (dimension of time$^{-1/2}$, with mean zero and delta-timecorrelated with unit variance). At time $s=0$, the particle has Maxwellian velocity distribution, with $\langle v\rangle_\text{eq} =0$. The idea is that  at times $s>0$, $F_s$ pushes the particle to move.  The mobility is the response function $R(\tau)$ entering in the expected velocity
\[
\langle v(t)\rangle_F = \int_0^t \id s \,R(t-s)\,F_s
\]	
It is showing how susceptible the particle is to the force $F_s$.  Here we can compute everything and find $R(\tau) = \frac 1{m}e^{-\gamma\tau}, \tau\geq0$. If $F_s\equiv F$ is contant in time,
\[
\lim_{t\uparrow \infty}\,\langle v(t)\rangle_F = \int_0^\infty \id s \,R(t-s)\,F = \frac 1{\gamma\,m}\, F
\]	
which means that the mobility ${\cal M} = 1/(\gamma m)$.\\
It is however physically and mathematically often useful to work in Fourier space.
One easily computes the Fourier transform,
\[
\tilde{R}(\nu) = \int \id t \,e^{i\nu t}\,R(t) = \frac 1{m}\frac 1{\gamma-i\nu}
\]
with imaginary part
\begin{equation}\label{ima}
\text{Im } \tilde{R}(\nu) = \frac 1{m}\frac{\nu}{\gamma^2+\nu^2}
\end{equation}	
On the other hand, without the forcing we have an equilibrium process, satisfying detailed balance with Maxwellian stationary distribution.  There, the time-correlation is $\langle v_t\,v_0\rangle_\text{eq} = \frac{k_BT}{m}\,e^{-\gamma |t|}$, such that its Fourier transform equals
\[
\tilde{G}(\nu) = \int\id t \,e^{i\nu t}\langle v_t\,v_0\rangle_\text{eq} = \frac{2k_BT}{m}\frac{\gamma}{\gamma^2 +\nu^2}
 \]
which implies the equality (with $\beta=1/(k_BT)$)
\begin{equation}\label{lan}
\text{Im } \tilde{R}(\nu) = \frac{\beta\nu}{2\gamma}\,\tilde{G}(\nu)
\end{equation}
The identity \eqref{lan} is an elementary example of the fluctuation--dissipation theorem. We will call it a fluctuation--dissipation relation (FDR) of the first kind.  In the present case it provides an easy example of the Sutherland-Einstein relation because the diffusion constant $\cal D$ is related to $\tilde G$: with $q_0=0$,
\[
\langle q^2_t\,|\,q_0=0\rangle_\text{eq} =  2 \int_0^t\id t_1\int_0^{t_1}\id t_2 \,\langle v_0\,v_{t_2}\rangle_\text{eq}
\]
so that
\begin{equation}\label{See}
{\cal D} := \lim_{t\uparrow\infty}\frac{\langle (q_t -q_0)^2\rangle_\text{eq}}{2t} = \frac 1{2} \tilde{G}(\nu=0)
	\end{equation}
Combining \eqref{ima}, \eqref{lan} and \eqref{See} we arrive indeed at
\begin{equation}\label{dm}
{\cal D} = \frac{k_BT}{\gamma\,m}= k_BT\,{\cal M},\qquad {\cal M} = \int_0^\infty \id \tau\,R(\tau) = \frac 1{\gamma \,m}
\end{equation}
Note that this relation is exact for the Langevin dynamics \eqref{lvin}, because of the linearity of the dynamics.\\
Looking back at the RC-circuit and \eqref{rclang1}, we see the same structure as in \eqref{lvin} with the identification $\gamma = 1/(RC), F_s={\cal E}/R$ and $m=C$.  In other words, it is the FDR of the second kind (also called, the Einstein relation between noise and friction) that ensures \eqref{dm}: if the factor in front of the noise in \eqref{lvin} would have been different, \eqref{dm} would not obtain.  In fact, also the opposite is true in the sense that the FDR of the second kind can be derived from linear response theory (relations like \eqref{dm}) around equilibrium for the bath. We will explain that in Section \ref{kind}. All of such relations depend on microscopic reversibility, which is derived from the dynamical reversibility of Hamiltonian dynamics in the microcanonical ensemble.  These things will become more clear as we proceed; see Section \ref{lodetb}.

\end{example}

\begin{example}\label{rw}[simple random walk]
Consider a dilute suspension of colloids being driven in a tube or channel with a rough and irregular inner surface and filled with some viscous fluid in equilibrium at temperature $T$. We suppose that the tube is spatially periodic in one dimension with cells of size $L$.  The driving is from a constant force $F$ (on the colloids) pushing them  say to the right. A picture of the situation in one cell (repeated periodically) is provided in Fig.~\ref{fig:tub}. We want to model that dynamics and transport with a biased continuous-time random walk on the one-dimensional lattice.  Each site $x$ corresponds to a cell.  Our mathematical model needs two parameters giving the transition rates to hop to the right, respectively to the left,
\[
k(x,x+1) = p,\qquad k(x,x-1) = q
\]
We think of a local force around $x$ making that possible, which is working on the walker to make the transition to the next cell.  That work is dissipated instantaneously into the thermal environment. The work done by the constant force over length $L$ is dissipated as Joule heating in the fluid. The corresponding change in entropy in the bath is thus $FL/T$.

\begin{figure}[!h]
	\centering
	\includegraphics[width=0.55\textwidth]{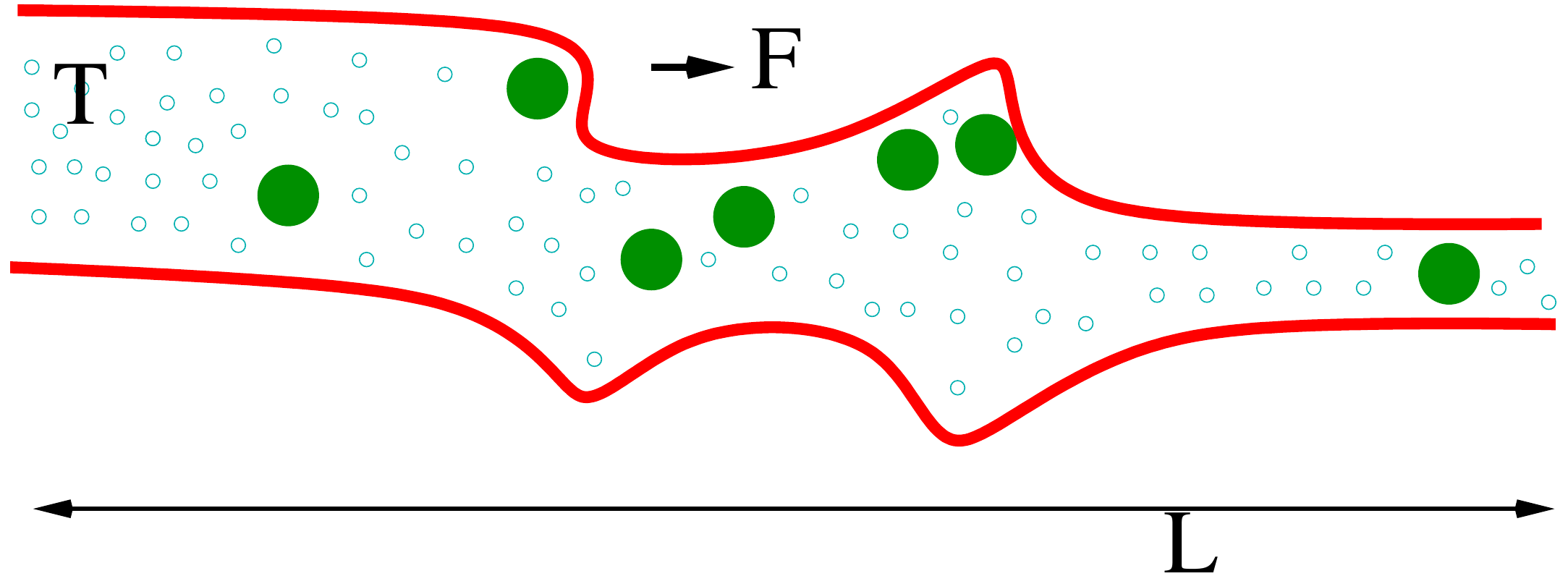}
	\caption{Flow in a rough channel with  periodically repeated cells of length $L$.  Colloids are pushed with force $F$ in a thermal bath at temperature $T$.  Trapping may occur causing the current to drop at larger $F$.}
	\label{fig:tub}
\end{figure}

 From the condition of local detailed balance (to be recalled in Section \ref{lodetb}) we put $p/q = \exp[FL/k_BT]$ which expresses that the ratio of forward to  backward rates is given by the entropy flux to the environment per $k_B$.    Writing $FL/k_BT = \epsilon$ we thus have
\begin{equation}\label{rwr}
k(x,x+1) = a(\epsilon)\,\exp[\epsilon/2],\qquad k(x,x-1) = a(\epsilon)\,\exp[-\epsilon/2]
\end{equation}
where we inserted a kinetic parameter $a(\epsilon) = \sqrt{pq} >0 $, possibly depending on the driving $F$, temperature $T$, cell length $L$ and other things such as the geometry of the channel/tube.  To say it differently we suppose the escape rate from each cell to be
\[
p+q = 2\,a(\epsilon)\,\cosh[\epsilon/2]
\]
It tells us how the average residence time $\sim 1/(p+q)$ in each cell of the channel depends on the force $F$.  Now let us see about the motion.\\
The current (flux per particle from cell to cell) obviously equals
\begin{equation}\label{ve}
\langle v\rangle_F= L\,( p - q) = L\,\frac{p - q}{p+q}\,(p+q) = 2L\,\frac{e^\epsilon - 1}{e^\epsilon+1}\,a(\epsilon)\,\cosh[\epsilon/2]
\end{equation}
Expanding around $F=0$ gives for the linear term
\[
\langle v\rangle_F = L\,a(0)\,\epsilon = a(0)\,\beta\,FL^2 
\]
Hence, the mobility  is $\cal M = \beta\,a(0)L^2$, the linear transport coefficient.
That is again an instance of the Sutherland-Einstein relation since the diffusion constant (without force, $\epsilon=0$) here equals ${\cal D} =  a(0)L^2 = \frac{k_BT}{F}\,\langle v\rangle_F = k_BT\,\cal M$.\\
Note that the expression \eqref{ve} is exact and can of course be evaluated to all orders in $\epsilon$.  The differential mobility $\id \langle v\rangle_F/\id F$ as function of $\epsilon$ clearly picks up the dependence of the escape rate $p+q$ on $\epsilon$.  In particular it is easy to see that this differential mobility can get negative at large enough values of $\epsilon$ when $p+q$ decreases with large $\epsilon$.  There is nothing surprising here, and we will see later how that conclusion can be turned into a constructive idea.
\end{example}

\begin{example}[Periodic potential]	\label{ser}
	Example \ref{lav} can be extended  to include a periodic potential.
Then, a force is added to the Langevin equation that derives from a
	periodic potential $U$, making (in three dimensions now)
	\begin{eqnarray}\label{diffpot}
		\dot{\vec{r}}_t &=& \vec{v}_t\nonumber\\
		m\,\dot{\vec{v}}_t &=& -\nabla U(\vec{r}_t)+ \vec{f} -\gamma m\, \vec{v}_t + \sqrt{2m\gamma \,k_BT}\,\vec{\xi}_t
	\end{eqnarray}
	where $\vec f$ is a constant force, to perturb the purely diffusive motion.
For the response, there is the mobility (matrix) function ${\cal M}(t)$, measuring the expected displacement of the particle:
	\[ M_{ij}(t) = \frac{1}{t}\left.\frac{\partial}{\partial f_j}\Big<({\vec r}_t-{\vec r}_0)_i\Big>_{f}\right|_{\vec{f}=0} \]
	The subscript $f$ in the average refers to the dynamics with the extra force $\vec{f}$, perturbing $-\nabla U(\vec{r}_t)
	\rightarrow -\nabla U(\vec{r}_t) + \vec{f}$. The mobility is the limit
	\begin{equation}\label{mobi}
M_{ij} = \lim_{t\to\infty}M_{ij}(t) 
 \end{equation}
	giving the linear change in the stationary velocity by the addition of a
	small constant force.  The subscripts give the components of the corresponding vectors.  \\
	The diffusion (matrix) function ${\cal D}(t)$ at finite time $t$ is defined as
	\begin{equation}\label{dij}
	D_{ij}(t) = \frac{1}{2t}\Big<(\vec{r}_t-\vec{r}_0)_i;(\vec{r}_t-\vec{r}_0)_j\Big>_\text{eq}
	\end{equation}
That is again in the equilibrium process, with $\vec f =0$. The right-hand side is the covariance: in general, for observables $A$ and $B$ we write
	\begin{equation}\label{trunc}
	\Big<A;B\Big>=\Big<AB\Big> -\Big<A\Big>\Big< B\Big>
	\end{equation}
The diffusion matrix is the limit
	\begin{equation}\label{diffi} D_{ij} = \lim_{t\to\infty}D_{ij}(t) \end{equation}
as we expect the (co)variance of the displacement of the particle to be linear in time $t\gg 1/\gamma$.\\

Exact computations are tedious now.  Yet we will see in Section \ref{gka} why (also for the dynamics \eqref{diffpot}) we have the standard Sutherland--Einstein relation $M_{ij} = D_{ij}/(k_BT)$.  Note however that in contrast with the case where $U=$ constant, the mobility no longer equals $\delta_{ij}/(\gamma m)$. For example, the mobility decreases with the amplitude of the conservative force as the particle
	needs to escape potential wells to have a non-zero velocity.
\end{example}

\section{Dynamical ensembles}\label{dynens}
Equilibrium statistical mechanics is centered around an object which is often called the Hamiltonian.  It specifies the interaction potential between the components. Given such an energy function on some effective scale of description, the ensemble gets fixed by specifying the constraints or by giving intensive variables such as temperature and chemical potential.  The resulting Boltzmann--Gibbs probability laws give the equilibrium distributions on configuration or phase space.  Under conditions like translation-invariance, they are solution of the Gibbs variational principle for a suitable free energy functional.\\
  There is no strict analogue for nonequilibrium systems, at least not reaching the power and the glory of the Gibbs formalism. While in some rare cases of nonequilibrium systems we have partial information about the stationary (single-time) distribution for a given dynamics, there is no overarching principle to specify it physically.  The reason is probably that kinetic (non-thermodynamic) features cannot be well represented (locally) at a fixed time.\\
  
   The situation appears to be more promising on trajectory space.  Such an option was already chosen in the work of Onsager and Machlup \cite{OM}, for Gaussian processes showing small fluctuations around hydrodynamical behavior for relaxation to equilibrium.  It was also the start of \cite{gibbs} for studying steady nonequilibrium.  We then want to find the physically correct (relative) weigths of trajectories, as traditionally given in terms of an action and a Lagrangian.  We see below how to construct the action for Markov processes.  Yet, and even more importantly, we hope to understand operationally what contributes to the action by using it.\\

The idea is to consider on the level of description of interest a family of possible (i.e., realizable) trajectories $\omega$.  They are realized by continuous time processes for systems in contact with possibly various but well-separated equilibrium reservoirs.  We open the time-window $[0,t]$ to write $\omega = (x_s, s\in [0,t])$ for a trajectory.  The ``state'' $x_s$ at time $s$ can be a many-body mesoscopic condition, e.g, giving the chemomechanical configuration of a collection of molecular motors or the positions of colloids or the displacements and velocities for a crystal of oscillators\footnote{We will write $x$ for a general state, possibly including many-body positions, velocities or spins.  We use $q$ or $\vec r$ when we explicitly address the positions of particles, and $v$ for velocities.}.  Most often, the space of trajectories (path-space) must be restricted mathematically to have some regularities and for sure, it is an infinite--dimensional space.  Yet, we ignore the mathematically more precise formulation, which is trivial enough, and we outline the formal structure only, choosing also for the simplest notation.  In that spirit we write
the probability of a trajectory as
\begin{equation}\label{aha}
\text{Prob}[\omega] = P[\omega] = e^{-{\cal A}(\omega)}\,P_\text{ref}[\omega]
\end{equation}
where the $P_\text{ref}= $ Prob$_\text{ref}$ denotes a reference ensemble (probability) and $\cal A$ is called the action. We obviously want to use that the action ${\cal A}$ as function of the trajectories, is (quasi-)local in spacetime.  E.g. for Markov processes, ${\cal A}$ will be given by a time-integral of single or double-time events. We did not specify here the initial conditions (at time $0$) but the idea is that we want ${\cal A}$ only to depend on the dynamics, not on the initial conditions\footnote{We do not consider in the present review the case of comparing two different initial conditions, or the relaxation from perturbing the stationary distribution as initial condition.  In such cases, a procedure following the Agarwal-method is possible, \cite{ag,njp}.}.  In other words, in \eqref{aha} we let $P(x_0) = P_\text{ref}(x_0)$, coninciding at time zero. Below we give examples to illustrate that structure; Section \ref{decoex} is devoted to it.  To start immediately however, we go back to Example \ref{rw}.

\begin{example}[simple random walk, continued]\label{contra}
What weight $P[\omega]$ to give to a trajectory $\omega$ of a continuous--time random walker? As in Example \ref{rw}, we take the transition rates $k(x,x+1) = p, k(x,x-1)=q$ on the one-dimensional lattice.  A trajectory has periods of waiting separated by jump times.  The waiting times are distributed exponentially with constant rate $p+q$, wherever the walker resides at that moment. It will contribute an overall factor. 
To concentrate on the jumping, we suppose the trajectory $\omega$ has $N_+$ steps forward and has $N_-$ steps backward during $[0,t]$.  Then,  
 \begin{eqnarray}\label{in}
P[\omega] &\propto& e^{-(p+q)t}\, p^{N_+(\omega)}\,q^{N_-(\omega)}  = e^{-(p+q)t}\, \left(\frac{p}{q}\right)^{(N_+(\omega)-N_-(\omega))/2}\,(pq)^{(N_+(\omega) + N_-(\omega))/2} \nonumber\\
&\propto&  e^{-(p+q)t}\, e^{\epsilon(N_+(\omega)-N_-(\omega))/2}\,a(\epsilon)^{N_+(\omega) + N_-(\omega)} \nonumber\\
&\propto&  e^{-(p+q)t}\, e^{\epsilon\,J(\omega)/2}\,a(\epsilon)^{N(\omega)} 
\end{eqnarray}  
where the second line takes the notation of Example \ref{rw} and, in the last line, $J(\omega)= N_+(\omega)-N_-(\omega)$ is the time-integrated (variable) current while $N(\omega) = N_+(\omega) + N_-(\omega)$  is the total number of (unoriented) jumps (dynamical activity). Hence, taking as reference the process with $\epsilon=0$  in \eqref{aha}, we have 
\begin{equation}\label{redec}
{\cal A}(\omega) = -N(\omega) \log \frac{a(\epsilon)}{a(0)} - \frac{\epsilon}{2}\,J(\omega)
\end{equation}
up to irrelevant (since constant) terms.  Let us see what we can learn from just that expression.  Take e.g. $-\log a(\epsilon)/a(0)\simeq \epsilon^2 $.  Then, for large $\epsilon$, trajectories $\omega$ having small $N(\omega)$ are preferred.  Therefore, as $\epsilon$ grows larger, the dynamical activity gets reduced and hence the current will also decrease. It will possibly die.    That is the same conclusion as from the considerations in Example \ref{rw}.
Trapping far-from-equilibrium can be induced by pushing too much;
see also \cite{zia,negheatcap}. On the other hand, for small $\epsilon$ (in linear order around zero bias) we can as well forget the influence of the dynamical activity and the linear response regime may be called purely dissipative: we could as well take 
\begin{equation}\label{mut}
P[\omega] \propto e^{\epsilon\,J(\omega)/2}\,P_\text{ref}[\omega]
\end{equation}
instead of \eqref{aha}--\eqref{redec}, when asking for linear response around the reference $\epsilon=0$.
\end{example}  

\subsection{Decomposition from time-symmetry}\label{tis}
In the generality in which we work at this point, there is only one but rather relevant symmetry transformation to decompose the action $\cal A$ in \eqref{aha}.  We consider the involution $\theta$ on trajectories $\omega$, by which
\begin{equation}\label{thet}
(\theta\omega)_s =\pi\omega_{t-s},\qquad s\in [0,t]
\end{equation}
The kinematical time-reversal $\pi$ is an involution on the state space which flips the odd degrees of freedom (such as velocities) present in the trajectory. We assume here that $\theta\omega$ is an allowed trajectory, whenever $\omega$ is (assumption of dynamical reversibility).  Note that we also time-reverse external (time-dependent) protocols, if any, in the same manner.\\
We now decompose the action according to that symmetry,
\begin{equation}\label{adec}
{\cal A} = D - \frac 1{2} S, \quad D := \frac 1{2}\left({\cal A} + {\cal A}\theta\right),\quad S :=  {\cal A}\theta - {\cal A}  
\end{equation}
The reason for the factor 1/2 in front of $S$ will become more clear later\footnote{It is the same $1/2$ as multiplying $J(\omega)$ in \eqref{redec}.}. The main point is that under the condition of local detailed balance (below), $S(\omega)$ is the change of entropy (per $k_B$) in the environment as caused and determined by the system trajectory $\omega$.  We therefore refer to $S$  (anti-symmetric under time-reversal $\theta$) as the entropic part.  The time-symmetric part $D$ is referred to as the frenetic part. Note that both $D$ and $S$ represent {\it excesses} with respect to the reference ensemble; they specify how entropic and frenetic parts change. A more informal observation may be that our Lagrangian approach \cite{gibbs,poincare} where we give weights to trajectories with the decomposition of the action $\cal A = D- S/2$ suggests to think of $D$ as the analogue of time-integrated kinetic energy and of $S$ as the analogue of time-integrated potential energy.  In that respect, the extensivity in time is only guaranteed for $D$.

\subsection{Examples}\label{decoex}

 The writing of \eqref{redec} already gives an example of the decomposition \eqref{adec}: the dynamical activity $N(\omega)$ is clearly time-symmetric, and the particle current $J(\omega)$ is time-antisymmetric.  Indeed, $\epsilon J(\omega)$ is the entropy flux per $k_B$ released in the viscous environment.  We give some other examples illustrating the decomposition.

\begin{example}[Markov jump processes]\label{mjp}
We denote the transition rate for a jump $x\rightarrow y$ by 
\begin{equation}\label{gfo}
k(x,y) = a(x,y)\,e^{s(x,y)/2}
\end{equation}
taking a parametrization with symmetric activity parameters
\begin{equation}
\label{apa}
a(x,y) = a(y,x) = \sqrt{k(x,y)k(y,x)}
\end{equation}
and antisymmetric driving
\[
s(x,y) = -s(y,x) = \log \frac{k(x,y)}{k(y,x)}
\]
 Under local detailed balance, also discussed in the next section, the $s(x,y)$ get interpreted as the (discrete) change of entropy per $k_B$ in the equilibrium bath with which energy, volume or particles are exchanged during the system transition $x\rightarrow y$.   Here, an environment is imagined consisting of spatially well-separated equilibrium baths, each with fast relaxation.  Trajectories are piecewise constant and they consist of ``waiting'' times and ``jumping'' events.  During the jump, the system exchanges ``stuff'' with one of the baths. Local detailed balance thus amounts here to being able to identify
	\begin{equation}
	\label{enp}
	S(\omega) = \sum_\tau s(x_{\tau^-},x_\tau)
	\end{equation}
with the path-wise total entropy flux (per $k_B$) in the environment.  In \eqref{enp} we sum over the jump times in the (system) trajectory $\omega = (x_\tau, 0\leq \tau\leq t)$ and $x_{\tau^-}$ is the state just before the jump to the state $x_\tau$ at time $\tau$.  In other words, we assume in such models that we can read the variable changes of the entropy in the reservoirs in terms of system trajectories.  Note of course that the path-wise entropy flux $ S(\omega) = -S(\theta\omega)$, is antisymmetric under time-reversal.\\
 Waiting between jump times takes a random time, exponentially distributed with the escape rate
\[
\xi(x) := \sum_y k(x,y)
\]
as parameter, when in state $x$.  The time-integrated escape rate equals
	\begin{equation}\label{esc}
	\text{Esc}(\omega) := \int_0^t\id s \,\xi(x_s)
	\end{equation}
	as function of the trajectory $\omega$ in $[0,t]$. Clearly, $\text{Esc}(\omega) = \text{Esc}(\theta \omega)$ is time-symmetric. There is also a second time-symmetric component in the jumping itself: the activated traffic can be measured from
	\begin{equation}\label{dd}
	\text{Act}(\omega) := \sum_s \log \frac{a(x_{s^-},x_s)}{a_0}
	\end{equation}
	where the sum is again over the jump times in $\omega$ and $a_0$ is a reference rate.\\

Let us finally turn to \eqref{adec}.	The frenesy  associated to the path $\omega$ is
\begin{eqnarray}
\label{fre}
D(\omega) &:=& \text{Esc}(\omega) - \text{Act}(\omega) \nonumber\\
&=& \int_0^t\id s\,\sum_y k(x_s,y)    - \sum_s \log a(x_{s^-},x_s)/a_0
\end{eqnarray}
\end{example}  
That makes the time-symmetric contribution in the decomposition \eqref{adec}. When $a(x,y)\equiv a$ is constant, Act$(\omega)$ is proportional to the dynamical activity (time-symmetric traffic, total number of jumps) over $[0,t]$.

\begin{example}[Overdamped diffusion]
We can take the diffusive limit of the previous example.  A Brownian particle has position ${\vec r}_t=(r_t(1),r_t(2),r_t(3)) \in {\mathbb R}^3$ with motion following
 \begin{equation}
 \dot{\vec r}_s = \chi \, \vec F({\vec r}_s) + \sqrt{2k_BT\,\chi}\,\xi_s,\qquad {\vec \xi}_s=\mbox{ standard white noise vector}\label{overd}
 \end{equation}
The mobility $\chi$ is a positive $3\times3-$matrix that for simplicity we choose not to depend on $q$ here. It implies that in the frenesy, only the escape rates will change when we change $F$ with respect to a reference choice.  We put
\[
\vec F(\vec r) = h\,\vec f(\vec r)  + \vec g(\vec r)
\]
where $\vec f$ and $\vec g$ are vector functions.  The constant $h$ is a parameter and $h=0$ gives the reference dynamics.
We want the excess frenesy and entropy flux per $k_B$ for $h\neq 0$, as defined from  \eqref{aha} and \eqref{adec}.  We refer to \cite{fren,fdr1,jmp2000} for detailed calculations.  Mathematical understanding follows from the Cameron-Martin and Girsanov theorems for the change of measure (via Radon-Nikodym derivative); cf \cite{girs}.  We can also remember the trick that ${\vec \xi}_s,s\in [0,t],$ is (formally) a stationary Gaussian process whose weights carry over to the trajectory via the quadractic form
 \begin{equation}\label{stc}
\frac 1{2}{\vec \xi}_s \cdot {\vec \xi}_s =  [\dot{\vec r}_s - \chi \, \vec F({\vec r}_s)]\cdot\frac 1{4k_BT\,\chi}\,[\dot{\vec r}_s - \chi \, \vec F({\vec r}_s)]
	\end{equation}
To obtain the action ${\cal A}$, that must be integrated over time $s\in [0,t]$ after which we must take the difference between the expressions for $\vec F = \vec g$ and for $\vec F=\vec g + h\vec f$.\\
At the same time we must be clear about the stochastic integration.  In \eqref{stc}, It\^o-integration must follow, which is not symmetric under time-reversal. It is useful to change to Stratonovich-integration therefore.  That rewriting uses the relation
\begin{equation}\label{is}
\int_0^t \vec G({\vec r}_s)\circ \id {\vec r}_s = \int_0^t \vec G({\vec r}_s)\,\id {\vec r}_s + k_BT\int_0^t(\chi \nabla) \cdot \vec G({\vec r}_s)\,\id s
\end{equation}
for general smooth functions $G$, 
that connects for \eqref{overd} the Stratonovich-integral (left-hand side) to the It\^o-integral (first term on the right-hand side).\\
Using that the Stratonovich-integral $\int_0^t \vec f({\vec r}_s)\circ \id {\vec r}_s$ is anti-symmetric under time-reversal, the result for \eqref{overd} is
\begin{eqnarray}\label{frda}
  D(\omega) &=& \frac{h^2\beta}{4}\int_0^t \id s\, \vec f\cdot \chi \vec f + \frac{h\beta}{2}\int_0^t \id s \,\vec f\cdot \chi\,\vec g + \frac{h}{2} \int_0^t \id s\, \chi\nabla\cdot \vec f\\
S(\omega) &=& h\,\beta\int_0^t \id {\vec r}_s \circ \vec f({\vec r}_s)\label{se}
\end{eqnarray}
in \eqref{adec}.
 Note that the highest order in the excess parameter $h$ appears in the frenetic part.  Indeed, frenesy will matter more at larger excesses.\\
 
  When ${\vec f}=\nabla V$ is conservative, then the second and third term in $D$ (the linear part of the frenesy) add up to become proportional to the time-integral of the backward generator $\cal L$ acting on $V$:
\[
\text{for } \vec f = \nabla V,\qquad \vec f\cdot \chi\,\vec g +  k_BT\chi\nabla\cdot \vec f=  {\cal L}V
\]
for the backward generator ${\cal L}u = \nabla u\cdot \chi \vec g + k_BT(\chi\nabla) \cdot \nabla u$ (on a functon $u$) of the reference dynamics.  On the other hand, the entropy flux \eqref{se} becomes a time-difference, $S(\omega) = h\beta[V({\vec r}_t)-V({\vec r}_0)]$.\\
    We can also specify to the case where $\vec g = -\nabla U$ and $\vec f$ being the nonconservative (or rotational) part of the force $\vec F$.  The reference dynamics ($h = 0$) satisfies the condition of detailed balance (time-reversibility). 
The excess frenesy \eqref{frda} now equals
\begin{equation}\label{fred}
  D(\omega) = \frac{h^2\beta}{4}\int_0^t \id s\, \vec f\cdot \chi\vec f - \frac{h\beta}{2}\int_0^t \id s \,\vec f\cdot \chi\nabla U + \frac{h}{2} \int_0^t \id s\, \chi\nabla\cdot \vec f
\end{equation}
The entropy flux per $k_B$ becomes time-extensive, being $\beta$ times the work done by the nonconservative
 force as given in \eqref{se}.  It is the Joule-heat divided by $k_BT$.
 \end{example}

 \begin{example}[Underdamped diffusion]\label{uex}
 The Langevin dynamics for a particle with mass $m$, position $q_s$ and velocity $v_s$ reads in one-dimensional notation as
 \begin{eqnarray}
  \dot{q}_s &=& v_s  \nonumber\\
  m\dot{v}_s &=& [F(q_s) + \epsilon\,f(q_s) - m\gamma v_s] + \sqrt{2{\cal D}}\,\xi_s\label{underd}
 \end{eqnarray}
 where we added a perturbation $f$ of strength $\epsilon$ to the reference force $F$.
Here, $\gamma$ is the constant friction and $\xi_s$ is standard white process, as always.  The strength ${\cal D} = m\gamma k_BT >0$ governs the variance of that noise.
 The action in \eqref{aha} is taken for force $F + \epsilon\,f$ with reference at  $\epsilon=0$.  The decomposition \eqref{adec} here employs the velocity-flip in the time-reversal.  The result gives, \cite{fdr2,col},
 \begin{eqnarray}\label{smo}
D(\omega) &=& \frac{\epsilon^2}{2{\cal D}}\int_0^t \id s f^2(q_s)
                          + \frac{\epsilon}{{\cal D}}\int_0^t \id s \,f(q_s)\,F(q_s) - m\,\frac{\epsilon}{{\cal D}} \int_0^t \id v_s \circ f(q_s)\\
  S(\omega) &=& \epsilon\beta\int_0^t \id s \,v_s \,f(q_s)\label{smos}
 \end{eqnarray}
 As before, $S$ equals the work done by the nonconservative force $f$, times $\beta$.  The frenesy $D$ represents the kinetics.  Note also that in the last two (linear in $\epsilon$) terms of \eqref{smo} we find the sum $-F(q_s)\id s  + m\id v_s = -m\gamma v_s \id s + \sqrt{2{\cal D}}\,\xi_s$ (multiplied with $\epsilon f(q_s)/{\cal D}$) representing the thermostating forces (friction plus noise) for the original dynamics.\\

 The same formul{\ae} hold for time-dependent forces.  Suppose we set (with $m=1=k_B$)
 \begin{eqnarray}\label{te}
 	\dot q_s &=& v_s\\
 	\dot v_s &=&  -\gamma\,v_s + F(q_s,\lambda_s) + \sqrt{2\gamma T} \,\xi_s
 \end{eqnarray}
 with a time-dependence in the force $F$  governed by an external protocol with parameter $\lambda_s$ at time $s$.
 The reference process for applying \eqref{aha} takes $F=0$.  The time-reversal must include the protocol;  we reverse it as $(\theta\lambda)_{s} =  \lambda_{t-s}$.\\
 We find for \eqref{adec},
 \[
 S(\omega) = {\cal A}(\theta\omega,\theta \lambda)  - {\cal A}(\omega,\lambda) =
 \frac 1{T} \int_0^t \id s\, v_s\,F(q_s,\lambda_s)
 \]
 which is the time-integrated power divided by temperature, instantly dissipated as Joule heat in the environment and given by \eqref{smos}.
 The frenesy $D = ({\cal A}\theta + {\cal A})/2$ as in \eqref{smo} equals
 \[
 D(\omega) =  \frac 1{4\gamma T}\,\int_0^t \id s\left[ F(q_s,\lambda_s)^2  -2\,\dot{v}_sF(q_s,\lambda_s)\right]
 \]
 where the first term refers to an escape rate and the second term (with Stratonovich integral) to the dynamical activity (having the acceleration $\dot{v}_s$). 
  \end{example}

 Other examples can be added; heat conduction networks are treated in \cite{heatcond,fren}.  More examples are collected in \cite{jmp2000,urna}.

\subsection{Local detailed balance}\label{lodetb}
The decomposition of Section \ref{tis} is especially useful when there is a physical meaning to $S$ and $D$ as excesses with respect to the reference ensemble. The previous examples have shown that $S$ and $D$ may indeed come with such a physical meaning. The time-symmetric part $D$ is the frenesy, collecting both the undirected traffic and the quiescence in the trajectory: too much waiting is punished when the escape rates are high and undirected traffic (also called, dynamical activity) is being stimulated when the time-symmetric activation part exceeds that of the reference ensemble. That was already clear in Example \ref{contra}.  We learn more about the role of $D$ in the following section.\\

Here we want to recall that in all previous examples, $S$ is the (excess) entropy flux (per $k_B$) with respect to the reference process.  That is not an accident.  It is an instance of what has been called {\it local detailed balance}, \cite{hal,derrida,leb,time,snak}.  The environment of the system consists of spatially separated equilibrium baths, each showing fast relaxation in the weak coupling with the system.  For Markov jump processes in Example \ref{mjp}, the $s(x,y)$ give the (discrete) change of entropy per $k_B$ in the equilibrium baths following an exchange of energy, volume or particles during the system transition $x\rightarrow y$.
For the other Markov diffusion examples, the relation between friction and noise has been chosen in exactly such a way as to satisfy local detailed balance and each time indeed the antisymmetric part $S$ is the time-integrated entropy flux measured in units of $k_B$\footnote{We use the letter $S$ (and not $\sigma$ or $\dot{S}$) to indicate the variable (path-wise, trajectory--dependent, random,...) time-integrated entropy flux, believing no confusion will arise here with the thermodynamic state function ``entropy.''  In fact, the entropy flux refers to a change in that entropy in the totality of equilibrium reservoirs making the environment of the system.}.  The ultimate reason is time-reversal invariance of the microscopic system (microscopic reversibility) in the microcanonical ensemble, which for return to equilibrium is expressed as the condition of detailed balance; see \cite{time}.  The main point is that in the microcanonical ensemble (giving equal probability to all phase-space points on the constant energy-volume-particle number surface), entropy itself is giving the weight of a condition: the microcanonical weight at a single time can be expressed with the Boltzmann writing of entropy, 
\[
k_B \,\log\text{ Prob}_\text{mc}[x] = \text{ entropy}(x)
\]  
Continuing to write Prob$_\text{mc}$ for the weight of a (physically coarse-grained) trajectory $\omega$ in the microcanonical ensemble, time-reversal invariance gives
\[
\text{Prob}_\text{mc}[\omega]
= \text{Prob}_\text{mc}[\theta\omega]\] 
Hence, for the conditional probabilities,
\[
\frac{\text{Prob}_\text{mc}[\omega\,|\omega_0 = x]}{\text{Prob}_\text{mc}[\theta\omega\,|\omega_t = y]} = \exp \frac 1{k_B}\{\text{entropy}(y)- \text{entropy}(x)\}
\]
The logarithm of the ratio of transition rates is given by the change of entropy. 
A particularly relevant reduced description is to take mesoscopic variables for a subsystem and a thermodynamic description for its environment (consisting of equilibrium baths).  Then, under weak coupling assumption, the above identities propagate on the level of the subsystem, \cite{time}.
In summary, working under the condition of local detailed balance implies that we assume that the time-antisymmetric term $S$ in
\begin{equation}\label{locd}
\frac{\text{Prob}[\omega]}{\text{Prob}[\theta\omega]} =  e^{S(\omega)}
\end{equation}
gives the time-integrated entropy flux per $k_B$ in excess with respect to the reference ensemble\footnote{Remember that the operation $\theta$ of time-reversal is supposed to work on all dynamical variables including the protocol.  Even though that protocol is fixed, its time-reversed version is to be taken in the denominator of \eqref{locd}.}.  To make sure, the probabilities ``Prob'' in \eqref{locd} really refer to the same process or ensemble, i.e., starting from the same initial distribution at time zero and generated with the same dynamics.\\
There are various reformulations of that, and also various more microscopic foundations which are known as fluctuation theorems \cite{GC,jar,rue,stat}; we refer to \cite{gibbs,jmp2000,time,crooks,verbi,poincare,2ndlaw} for some of the original papers making the connection between the source term of time-reversal breaking and entropy.\\  As a final word of warning, we emphasize that not in all physical cases local detailed balance needs to be true.  For example, if a system is directly coupled to a nonequilibrium bath or if the coupling with or between equilibrium reservoirs is too large, local detailed balance will fail. We give two examples (and their response relations) in Section \ref{nl}.
	
\section{Response relations}\label{res}

\begin{figure}[!h]
 	\centering
 	\includegraphics[width=0.55\textwidth]{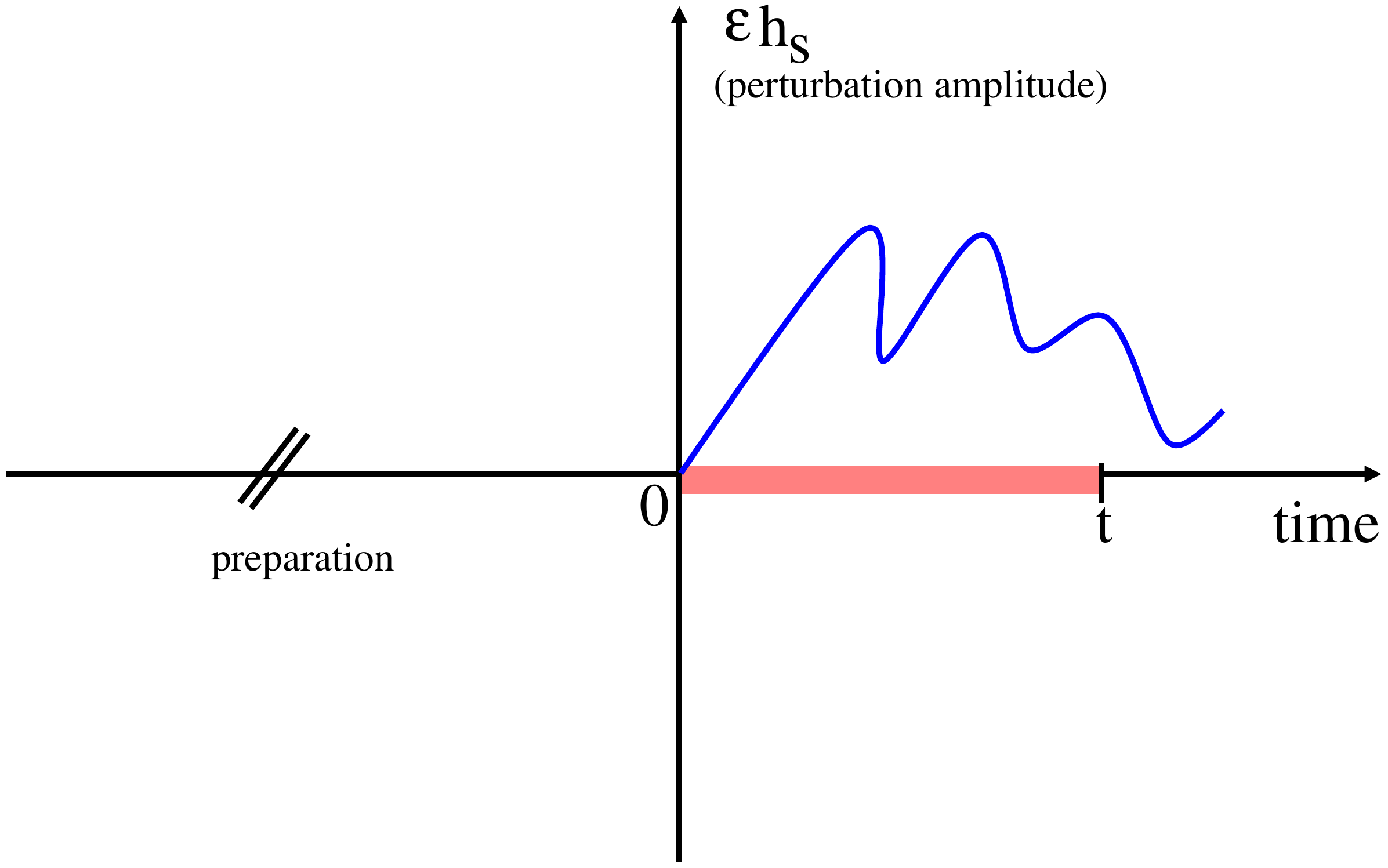}
 	\caption{The setup of response theory. Response is monitored at times $[0,t]$ from a stimulus in that same time-window and depending on the initial preparation (before time zero).  The (small) amplitude of the perturbation may vary in time.}
 	\label{fig:res}
 \end{figure}

 We come to the questions of Section \ref{x}. Recall the situation pictured in Fig.~\ref{fig:res}.\\

  In the present section we use dynamical ensembles to obtain response relations.  That is a different approach than from imitating classically the formalities of quantum mechanics and its perturbation expansions, \cite{kub,green}.  The reference ensemble is the original, unperturbed ensemble with reference probability Prob$_\text{ref}= P_0$.   The stimulus modifies the dynamical ensemble to the perturbed one, Prob $= P_\epsilon$, where we use $\epsilon$ to indicate the order of the spacetime amplitude of the perturbation.  Following \eqref{aha}--\eqref{adec}, we thus write
\begin{equation}\label{spe}
\text{Prob}[\omega] = e^{-\Delta D(\omega) + \frac 1{2} \Delta S(\omega)}\,\text{Prob}_\text{ref}[\omega]
\end{equation}
where we emphasize via ``$\Delta$'' that the perturbed ensemble shows changes in entropy flux and frenesy as caused by the perturbation over time $[0,t]$. The perturbed ensemble at time zero starts from the same distribution as the original reference.  It is important here to recall that the separation between time-symmetric (frenesy $D$) and time-antisymmetric (entropy flux $S$) contributions is obtained via the time-reversal operation ($\theta$ in \eqref{thet}) which should include the perturbation protocol; i.e., we also reverse the time-dependence in the perturbation, cf. the dynamics \eqref{te}.\\

Let us take an observable, i.e., a function $O$ of the trajectory $\omega$, always in the window $[0,t]$. By \eqref{spe}, its average in the perturbed ensemble is\footnote{Integration over trajectories is a mathematical subject we are not touching here; in line with a more common physics notation, we can also write $ P_0(\id\omega) =  P_0(\omega)\,\id\omega$.}
\begin{equation}\label{peens}
\langle O\rangle_\epsilon = \int P_0(\id\omega)\,O(\omega)\,e^{-{\cal A}(\omega)} = \langle O\, e^{-\Delta D + \frac 1{2} \Delta S}\rangle_0
\end{equation}
To indicate the strength of the perturbation we sometimes write a subscript on the expectations $\langle\cdot\rangle = \langle\cdot\rangle_\epsilon$.  Remember that the right-hand side is an average in the reference ensemble. In other words,
\begin{equation}\label{peensb}
\langle O\rangle_\epsilon - \langle O\rangle_0=  \left<O\, \left[e^{-\Delta D + \frac 1{2} \Delta S} - 1\right]\right>_0
\end{equation}
To show the order of perturbation we write $\Delta D = \epsilon\,D'_0 + \frac{\epsilon^2}{2}D''_0 + \ldots,\quad \Delta S = \epsilon\,S'_0 + \frac{\epsilon^2}{2}S''_0 + \ldots$ where the primes denote derivatives with respect to $\epsilon$ and $\epsilon$ is the strength (overall amplitude) of the considered perturbation.  The rest is straight; we expand the exponential in \eqref{peens} which to second order in $\epsilon$ turns into
\begin{eqnarray}\label{wonders}
\langle O\rangle_\epsilon - \langle O\rangle_0 &=& \epsilon\left\langle O \left[-D'_0 + \frac 1{2} S'_0 \right]\right\rangle_0 \\
&& +\;\frac{\epsilon^2}{2}\, \left\langle O \left[-D''_0 + \frac 1{2} S''_0  + (D'_0)^2 + \frac 1{4} (S'_0)^2 - D'_0S'_0\right]\right\rangle_0\nonumber
\end{eqnarray}
  For time-dependent perturbations the same logic applies. For what is next, we divide in various cases to estimate the relevant terms in the decomposition.  We start with the linear response around equilibrium.

\subsection{Linear response around equilibrium}\label{are}

Linear response takes the first order in the response formula of \eqref{wonders}.  We get
\begin{equation}\label{linwon}
\langle O\rangle_\epsilon - \langle O\rangle_0 = \epsilon\left\langle O \left[-D'_0 + \frac 1{2} S'_0 \right]\right\rangle_0 
\end{equation}
Remember that $D'_0,S'_0$ are the first derivatives evaluated at $\epsilon =0$. Note that, if we would have $O(\omega) = g(x_0)$, only depending on the initial time as some arbitrary function $g$, then $\left\langle O \left[-D'_0 + \frac 1{2} S'_0 \right]\right\rangle_0 =0$ by the normalization $\langle g(x_0)\, \exp[-{\cal A}]\rangle_0 =1$, as it should because $\langle g(x_0)\rangle_\epsilon = \langle g(x_0)\rangle_0$ and $\cal A$ only depends on the dynamics.  Such arguments take care of causality, that the response to later perturbations must equal zero.\\

Let us now focus on reference processes which are time-reversal invariant: $\langle O(\theta\omega)\rangle_0 = \langle O(\omega)\rangle_0$ or $P_0(\omega) = P_0(\theta\omega)$.  That is the case of reference equilibria, where we write expectations $\langle\cdot\rangle_{\text{eq}} =\langle\cdot\rangle_\text{ref} = \langle\cdot\rangle_0$. Linear response around equilibrium has been developed since the 1950's into a systematic theory, \cite{ku,chan,ba,ma,zwa,kub,green}.  We refer to \cite{spohn} for a review in the case of interacting particle systems.\\
Suppose first that the observable is odd under time-reversal, $O(\theta\omega) = -O(\omega)$: then, $\langle D'_0\, O \rangle_\text{eq} =0$ because $D'_0(\theta\omega) =D'_0(\omega)$ is symmetric and hence $D'_0\,O$ is antisymmetric and vanishes in equilibrium.  As a consequence, only the entropic contribution remains in the linear response formula \eqref{linwon}:  when $O\theta=-O$,
\begin{equation}\label{eqlin}
\langle O \rangle_\epsilon  =  \frac{\epsilon}{2}\,\left\langle O\, S'_0\right\rangle_\text{eq}
\end{equation}  
which is nonzero because $S'_0(\theta\omega) = - S'_0(\omega)$ is also anti-symmetric. This formula is generally true for linear response around equilibrium for odd observables and will be applied for state functions (as in the Kubo formula next) and for currents (in the Green-Kubo relations further down).   It is physically useful because of the ready interpretation of $S'_0$ as the (linear) excess (time-integrated) entropy flux due to the perturbation, following local detailed balance.  In particular we have that always
\begin{equation}\label{sf}
\langle S'_0\rangle_\epsilon =  \frac{\epsilon}{2}\,\left\langle (S'_0)^2\right\rangle_\text{eq} \geq 0
\end{equation}
which says that in linear order the expected dissipation in the perturbed condition is always nonnegative and equals the equilibrium variance of that flux.  That explains somewhat the origin of the terminology for the relation \eqref{eqlin} as fluctuation--dissipation relation (of the first kind).  The reason why the time-symmetric frenesy $D'_0$ is unseen in the linear response of (antisymmetric) currents $J$ is that, to linear order in $\epsilon$, field-reversal is equivalent with time-reversal.  To say it with a formula, we can as well use \eqref{mut} in linear response:
\[
\langle J \rangle_{-\epsilon} \simeq  \langle J\theta\rangle_\epsilon = - \langle J \rangle_\epsilon
\]
Such equivalence is of course not true in general farther away from equilibrium, except in very rare cases.  For such a rare case we refer to \cite{kpz} for an application of linear response in the context of directed polymers relevant for the fluctuations following the Kardar-Parisi-Zhang equation.\\

Secondly, when the observable $O$ is time-symmetric, like $D'_0$ itself, then we need the correlation between $O$ and the frenesy:
when $O\theta=O$,
\begin{equation}\label{eqlinn}
\langle O \rangle_\epsilon -  \langle O \rangle_\text{eq}  =  \epsilon\,\left\langle O\, D'_0\right\rangle_\text{eq}
\end{equation}  
That is interesting for currents which are even under time-reversal as happens for the momentum current (e.g. generated by shear).  Another example for jump processes is the number $N(\omega)$ of jumps (dynamical activity) in $[0,t]$ as in \eqref{in}--\eqref{redec}. Here we have that always
\[
\langle D'_0 \rangle_\epsilon -  \langle D'_0 \rangle_\text{eq}  =  \epsilon\,\left\langle (D'_0)^2\right\rangle_\text{eq} \geq 0
\]
which is the analogue of \eqref{sf}.
For example, again looking at \eqref{redec}, the expected change in the number of steps $\langle N\rangle_\epsilon - \langle N\rangle_\text{eq}$ for a random walker always has the same sign as $a'(0)$ for small $\epsilon$.

\subsubsection{Kubo formula}\label{kfr}
We can specify the result \eqref{eqlin} further by taking $O(\omega) = f(x_t) - f(\pi\,x_0)$ for a function $f$ on states.  We then go for single-time observations.  Remember here that $\pi$ is the kinematical time-reversal (like flipping the velocities if any).  In that case, the left-hand side says
\[
\langle O \rangle_\epsilon = \langle f(x_t) - f(\pi x_0)\rangle_\epsilon = \langle f(x_t)\rangle_\epsilon  - \langle f\rangle_\text{eq}
\]
where the last equality uses that we have equilibrium (full time-reversal invariance) at time zero.  For the right-hand side of \eqref{eqlin},
\[
\left\langle O \,S'_0\right\rangle_\text{eq} =   2\left\langle f(x_t) \,S'_0\right\rangle_\text{eq}
\]
where we used that $\left\langle f(\pi\,x_0) \,S'_0\right\rangle_\text{eq} = -\left\langle f(x_t) \,S'_0\right\rangle_\text{eq}$.  Hence, in linear response around equilibrium,
\begin{equation}\label{kub}
\langle f(x_t)\rangle_\epsilon  - \langle f\rangle_\text{eq} = \epsilon\,\left\langle f(x_t) \,S'_0\right\rangle_\text{eq} 
\end{equation}
for all functions $f$.  This response relation has followed straightforwardly from the assumption of time-reversal invariance in the equilibrium (reference) ensemble, where $S= \epsilon\,S_0' + O(\epsilon^2)$ is the antisymmetric part in the action ${\cal A}$ of \eqref{adec} or of \eqref{spe} under time-reversal,
following \eqref{locd}.
The final step for recognizing the Kubo formula in \eqref{kub} thus comes from the physical interpretation of \eqref{locd}: from local detailed balance, $S(\omega)$ is the entropy flux (per $k_B$) into the equilibrium environment due to the perturbation as seen from the system trajectory $\omega$.  We have announced that in Section \ref{lodetb} after giving the Markov dynamics examples in  
Section \ref{decoex}.  That implies for example that if the perturbation is opening a new energy exchange with potential $V(x) = V(q)$ (depending on positions $q$) and time dependent amplitude $\epsilon\,h_s, s\in [0,t]$, then the change of energy in the environment is
\[
\Delta E = \epsilon h_tV(x_t) - \epsilon h_0V(x_0)
\]
while the work done on the thermal bath equals
\[
W = \epsilon\,\int \id s \,\dot{h}_s V(x_s)
\]
Therefore, applying Clausius relation to the thermal equilibrium reservoir, the entropy change in the environment per $k_B$ is 
\begin{equation}\label{ass}
S(\omega) =\frac 1{k_BT} [\Delta E - W]=  \epsilon h_tV(x_t) - \epsilon h_0V(x_0) - \epsilon\,\int \id s\,\dot{h}_s V(x_s)
\end{equation}
as a function of the system position-trajectory $q_s, s\leq t$.  The correlation in \eqref{kub} becomes
\begin{eqnarray}\label{kubo}
\left\langle f(x_t) \,S'_0\right\rangle_\text{eq} &=& \left\langle f(x_t) \,[h_tV(q_t) - h_0V(x_0) - \,\int \id s \,\dot{h}_s V(x_s)]\right\rangle_\text{eq} \nonumber\\
&=&
 \frac  1{k_BT}\,\int \id s\, h_s \frac{\id }{\id s}\left\langle f(x_t) \,V(x_s)\right\rangle_\text{eq}
\end{eqnarray}
Concluding, we find that the linear response function for observing $f$ at time $t$ with perturbation of the energy  $E\rightarrow E-\epsilon\,h_s V(X_s)$ at time $s$
equals
\begin{equation}\label{kubs}
\frac{\delta \langle f(x_t)\rangle}{\delta (\epsilon h_s)}_{|_{\epsilon=0}} = R_{fV}(t,s) = \beta\, \frac{\id }{\id s}\left\langle f(x_t) \,V(x_s)\right\rangle_\text{eq}
\end{equation}
which is the Kubo formula \cite{kub,njp}.  Very little algebra has been used to derive it; yet the derivation is physically cogent.\\
 There are of course other possibilities for the entropy flux \eqref{ass}.  For example, in an underdamped dynamics we may have
 \begin{equation}\label{asst}
 S(\omega) =\frac {\epsilon}{k_BT} \int \id s\,h_s\,v_s\,\frac{\id V}{\id q_s} 
 \end{equation}
 as the time-integrated dissipated power over thermal energy (instead of \eqref{ass}).  That leads however to exactly the same Kubo formula \eqref{kubs} when using that $\dot{q}_s = v_s$.
 \\

We emphasize that we have not used any specific dynamical evolution except for the assumptions \eqref{ass} or \eqref{asst} which are physically motivated and readily derived for all the mesoscopics with a clear interpretation of entropy flux such as in all examples of the paper.   It means that we imagine the nonequilibrium process to proceed as if locally each transition or each local change in the state (in energy, particle number, volume or momentum) is in contact with one well-defined equilibrium reservoir, for which the condition of detailed balance \eqref{locd} applies.

\subsubsection{Green-Kubo and Sutherland--Einstein formula}\label{gka}
Another instance of \eqref{eqlin} is to take $O(\omega)=J_i(\omega)$, an antisymmetric current of some type $i$ (particles, energy, mass,...).  We follow again the condition of local detailed balance (Section \ref{lodetb}) whereby, when thermodynamic forces $\epsilon\,  F_k$ are exerted, then $S = \epsilon\sum_k J_k(\omega) \,F_k$. As a consequence we have
\begin{equation}\label{gkr}
\langle J_i \rangle_\epsilon =  \frac{\epsilon}{2}\,\left\langle J_i\, J_k\right\rangle_\text{eq}\,F_k
\end{equation}  
which are the Green-Kubo relations announced in \eqref{greku}.  A detailed modeling of some thermo-electric phenomena as introduced along the cartoon of Fig.~\ref{fig:the} and following local detailed balance is exposed in \cite{wier}.\\
Green-Kubo relations connect transport coefficients with fluctuation properties $\left\langle J_i\, J_k\right\rangle_\text{eq}$ in the equilibrium system.  Quite generally, in equilibrium, the latter can be rewritten as Helfand moments, mean square deviations in generalized displacements. In other words, the Green-Kubo relation gives so called Einstein-Kubo-Helfand expressions for transport coefficients.  The response can then be calculated as a (generalized) diffusion constant, \cite{hel,gasp}.  The best known example is the Sutherland--Einstein relation as discussed in Section \ref{exam}, Examples \ref{lav} and  \ref{ser}.  We can now see its derivation more generally.\\

The Sutherland--Einstein relation tells that diffusion matrix and mobility are proportional,
\begin{equation}\label{einstein} 
M_{ij} = \frac{1}{k_BT}{\cal D}_{ij} 
\end{equation}
where we use the notation from Example \ref{ser}.  To understand its origin, we can use \eqref{gkr} or directly derive it from \eqref{eqlin}.  Taking a colloid suspended in a fluid at rest, we apply an external field ${\cal E}$.  The entropy flux per $k_B$ caused by dissipating the work done by the force is equal to
\[
S(\omega) = \frac 1{k_BT}\,{\cal E}\cdot (\vec{r}_t-\vec{r}_0) 
\]
As observable $O$ we take the displacement $\vec{r}_t-\vec{r}_0$ and apply \eqref{eqlin}:
\[
\langle \vec{r}_t-\vec{r}_0\rangle^{\cal E} = \frac 1{2} \langle (\vec{r}_t-\vec{r}_0)\,\frac 1{k_BT}\,{\cal E}\cdot (\vec{r}_t-\vec{r}_0)\rangle_\text{eq}
\]
Dividing by time $t$ and taking derivatives with respect to the force components ${\cal E}(i)$, we see that \eqref{dij} appears in the right-hand side and hence \eqref{einstein} is obtained.

\begin{remark} There remains often the question whether all this and all that are restricted to stochastic dynamics.  The correct answer starts from noting that in the correct (e.g. weak coupling) regime of reduced descriptions the {\it correct} dynamics is of course stochastic when considering the reduced trajectories only.  Obviously, the {\it same} result will be reached when doing the Hamiltonian dynamics in the bigger microscopic system, when the reduced dynamics made any sense to start with.  Deviations will be observable (experimentally) due to realistic couplings, finite time-scale differences or absence of thermodynamic limits etc. In other words, whenever we see $\langle \cdot\rangle_\text{eq}$ we better take an average over the microcanonical ensemble with suitable constraints of energy, volume, etc....  when we can.   Another consideration is the effectiveness of simulations which may be better for deterministic dynamics.  Note however that at any rate we must somehow circumvent the van Kampen objection in Remark \ref{vk} and take a statistical approach, meaning to observe the appropriate physically coarse-grained observables.  
	\end{remark}

\subsubsection{Fluctuation--dissipation relations of first and second kind}\label{kind}
The terminology of fluctuation--dissipation relations (FDRs) is not always very precise.  For better systematics, results on response relations in the linear regime around equilibrium are called FDRs of the first kind. E.g., we call the Kubo formula \eqref{kubs} an FDR of the first kind.  Often one focuses on the relation between mobility and diffusion.  As we explained just above and have illustrated in Section \ref{exam} with two examples, particle diffusion is related with mobility, obtained from measuring the induced velocity after applying an external field (Sutherland--Einstein relation).  There is however also an FDR of the second kind, called Einstein relation.\\
To avoid misunderstandings we speak about (1) the Sutherland-Einstein relation when meaning the linear response formula for mobility in terms of the diffusion, and (2) the Einstein relation when dealing with the connection between friction and noise.  The Sutherland-Einstein relation is a direct application of linear response theory around equilibrium, meaning the ensemble of Kubo and Green--Kubo relations.  The Johnson-Nyquist relation was among the first examples of an FDR of the second kind; see \eqref{nyq}.\\

In the set-up of Example \ref{lav}, the FDR of the first kind and of the second kind are about identical (hence the possible confusion of terminology).  In general indeed, the noise amplitude need not be equal to the (long-time) diffusion constant\footnote{For Markov diffusions, it is always related to the short-time mean square displacement though.}. There are however important connections between the FDR of the first and of the second kind, the glue being provided by the condition of local detailed balance of Section \ref{lodetb} and the source being time-reversal invariance.  To summarize the situation: imposing local detailed balance in the set-up of Markov dynamics as in Section \ref{decoex} implies an FDR of the second kind, which in turn implies a standard FDR of the first kind around equilibrium. Alternatively, imposing FDR of the first kind for the thermal equilibrium environment of a system, implies local detailed balance and the FDR of the second type for the system weakly coupled to that environment.  The various FDRs are, in other words, not equivalent but still strongly related.  Here we elaborate on the derivation and the nature of the Einstein relation (FDR of the second kind) which we first encountered as the Johnson-Nyquist relation in \eqref{nyq}.\\

The Einstein relation is best known from the relation between friction and the noise amplitudes for Brownian particles.  The physical origin of friction and noise is indeed one and the same.  Good experience has taught us that a colloid suspended in and moving through an environment of many much faster and smaller particles experiences friction and statistical fluctuations at the same time.  The relation with the FDRs of the first kind derives from the fact that the motion of a probe (e.g. a colloid) in a thermal bath can be considered as a stimulus there. It is a time-dependent perturbation on the equilibrium bath.  That environment responds and that feeds back to the probe motion, making friction and noise.\\ 
To be more specific, let us consider a probe trajectory $(Y_s, s\leq t)$ up to time $t$ as a perturbation from the case where the probe has always been at rest at its present position $Y_t$.  For the equilibrium bath coupled to the probe, that means (for \eqref{aha}) that we have the reference ensemble for the bath having the probe at rest (at position $Y_t$ at time $t$) and the perturbed bath ensemble where the probe moves away from $Y_t$ for time $s<t$:
\begin{equation}\label{app}
P(\omega|Y_s, s\leq t) = \exp [-D(\omega) + \frac 1{2} S(\omega)]  \;  P(\omega|Y_s=Y_t, \text{ for all } s\leq t) 
\end{equation}
The $P(\omega|Y_s, s\leq t)$ is the probability of a bath-trajectory $\omega$ conditioned on a(n arbitrary) probe trajectory $(Y_s)^t$, while the probability in the right hand-side is the reference probability on bath trajectories.\\
The difference between the two ensembles originates physically from the coupling between probe and bath.  
We assume for simplicity that the probe position $Y$ only enters via an interaction potential $U(Y,q) = 
\sum_{i=1}^N u(Y-q(i))$ with the various ($N$) bath particles at positions $q(i)$.  At time $s\leq t$, the force of the probe on a bath particle (with generic position $q$)  is thus of the form\footnote{We prefer to use one-dimensional notation for simplicity only.}
\[
u'(Y_s-q) = u'(Y_t-q) + (Y_s-Y_t)\,u''(Y_t-q) =  u'(Y_t-q) + h_s\,V'(q)
\] 
to linear order in $Y_s-Y_t$.
In the last equality, we rewrote the force to make the link with the notation of the response theory above:  $h_s = Y_s-Y_t$ is a time-dependent amplitude and $V(q)= u'(Y_t -q)$  for $s\leq t$ and at fixed $Y_t$.  In other words, the effect of the probe motion on the bath is to provide a time-dependent perturbation with potential $V$, much the same way as treated in the Kubo formula of Section \ref{kfr}.\\
Let us next find the relevant bath observable for which we need to see the influence of that perturbation. That has of course everything to do with the probe dynamics:  the force of each bath-particle on the probe (all at time $t$) is
\begin{eqnarray}\label{fobi}
-u'(Y_t-q_t) &=&  - \int\id \omega \,
P(\omega|Y_s, s\leq t)\,u'(Y_t - \omega_t) + \zeta_t,\nonumber\\
\zeta_t  &:=& \int\id \omega \,
P(\omega|Y_s, s\leq t)\,u'(Y_t - \omega_t) - u'(Y_t- q_t)  
\end{eqnarray}
where the fluctuation term $\zeta_t$ has mean zero for every probe trajectory $(Y_s, s\leq t)$.
For $\int\id \omega \,
P(\omega|Y_s, s\leq t)\,u'(Y_t - \omega_t)$ we use the Kubo formula \eqref{kubs}:  
\begin{eqnarray}\label{fob}
-u'(Y_t-q_t) &=&  - \langle u'(Y_t-\omega_t)\rangle^{Y_t} -\left\langle S(\omega, (Y_s)^t)\,;\,u'(Y_t-\omega_t)\right\rangle^{Y_t}+ \zeta_t\nonumber\\ 
\left\langle S(\omega, (Y_s)^t)\,;\,u'(Y_t-\omega_t)\right\rangle^{Y_t} &=& \beta \int^t_{-\infty} \id s \,\dot{Y}_s \,\left\langle u' (Y_t - \omega_s)\,;\, u'(Y_t-\omega_t)\right\rangle^{Y_t} 
\end{eqnarray}
where the average $\langle\cdot\rangle^{Y_t}$ with the probe at rest in $Y_t$ is taken over the stationary bath-particles.%\footnote{We use the notation $\langle A\,;B\rangle =\langle AB\rangle - \langle A\rangle\langle B\rangle$ to denote the covariance.}.
The identity \eqref{fob} follows from the entropy flux as time-integrated dissipated power by the probe on the bath,
\[
S(\omega, (Y_s, s\leq t)) =   \beta\,\int^t_{-\infty} \id s \,\frac{\id}{\id s}(Y_s-Y_t)  \,V(q_s) 
\]
Since the bath is supposed in thermal equilibrium, we indeed only need the entropic contribution for calculating the response.
The last term in the first line of  \eqref{fob} is the  noise introduced in \eqref{fobi} and given in zero order as
\[
\zeta_t^0(Y_t) = \langle u'(Y_t-q_t)\rangle^{Y_t} -u'(Y_t-q_t), \quad  \langle \zeta_t^0(Y_t) \rangle^{Y_t}=0
\]
while the time-correlations are
\begin{equation}\label{sta}
\langle \zeta_t^0(Y_t)\zeta_s^0(Y_t) \rangle^{Y_t} = \langle u' (Y_t-q_s)\, ;\, u'(Y_t-q_t)\rangle^{Y_t} 
\end{equation}
As a summary, the induced force on the probe at time $t$ is
\begin{equation}
-\langle u'(Y_t- x)\rangle^{Y_t} - \beta \int^t_{-\infty} \id s \,\dot{Y}_s \,\left\langle u' (Y_t - q_s)\,;\, u'(Y_t-q_t)\right\rangle^{Y_t} +  \zeta_t^0(Y_t)
\end{equation}
The first term is a systematic force on the probe.  The second term is the friction and the third term is the noise in linear order around the equilibrium bath, satisfying \eqref{sta}. We conclude therefore that it is the entropic term in the action that produces the Einstein relation between the noise kernel and the friction memory.  We do not elaborate here on the collective effect of the large $N$ number of bath particles which would have to be combined with a weak coupling limit; cf. the van Hove limit \cite{chan,zwa,vanh}.  That would simplify the expressions more, producing e.g. Gaussian white noise and a deltacorrelated-memory kernel in the friction.

\subsection{Linear response around nonequilibrium}
We move to the situation where the system's condition was prepared in steady nonequilibrium (until time zero). Note that we do not require a close-to-equilibrium regime, the perturbation is small but the reference condition can be far out-of-equilibrium.  The formalism applies generally, but for the interpretation we stick to the regime where we have local detailed balance; see Section \ref{lodetb}.  We still have \eqref{linwon} for perturbations around nonequilibrium, but we must include the frenetic contribution even in linear order.  Taking as observable $O(\omega) = f(x_t)$ a function of the state $x_t$ at time $t$, we get
\begin{eqnarray}
\langle f(x_t) \rangle_\epsilon - \langle f(x_t) \rangle_0 &=& \frac{\epsilon}{2} \langle f(x_t)\,  S'_0(\omega) \rangle_0 - \epsilon\langle f(x_t)\,D'_0(\omega) \rangle_0\nonumber\\ 
&=& \epsilon\,\langle f(x_t)\,  S'_0(\omega) \rangle_0 - \epsilon\left< \,f(x_t)\,\left[ D'_0(\omega) + S'_0(\omega) /2\right]\,\right>_0\label{kuku}
\end{eqnarray}
The last line has its first term on the right-hand side giving the Kubo formula \eqref{kubs} for linear response around equilibrium.  Indeed, time-reversal invariance in equilibrium implies $\langle \,f(x_t)\,\left[ D'_0(\omega) + S'_0(\omega) /2\right]\,\rangle_\text{eq} = \langle \,f(\pi x_0)\,\left[ D'_0(\omega) - S'_0(\omega) /2\right]\,\rangle_\text{eq} = 0$ because of the normalization$ \langle e^{-{\cal A}}\rangle_\text{eq}=1$ for whatever initial condition.
The correction to the linear response in equilibrium is (obviously) additive.  We can massage it into a multiplicative correction with respect to the Kubo formula \eqref{kubo} or \eqref{kubs}  by writing \eqref{kuku} as
\[ %\]\begin{equation}
\langle f(x_t) \rangle_\epsilon - \langle f(x_t) \rangle_0  =\epsilon\,\beta\left(1  -  \frac{\langle \,f(x_t)\,\left[ D'_0(\omega) + S'_0(\omega) /2\right]\,\rangle_0}{\langle f(x_t)\,  S'_0(\omega) \rangle_0}\right)\,\int_0^t\id s \,h_s\frac{\id}{\id s}\,\langle f(x_t)\, V(x_s) \rangle_0 \label{kukul}
\] %end{equation}
we get a prefactor 
\[
\beta_\text{eff} =  \left(1  -  \frac{\langle \,f(x_t)\,\left[ D'_0(\omega) + S'_0(\omega) /2\right]\,\rangle_0}{\langle f(x_t)\,  S'_0(\omega) \rangle_0}\right) \,\beta 
\] which may be called an {\it effective} inverse temperature when compared to \eqref{kubo}--\eqref{kubs}.  That is one way for an effective temperature to appear, obviously depending on the observable $f$; see e.g.~\cite{cug,cug2,pug}.  For example, if $\langle f(x_t)\,D'_0(\omega)\rangle_0 \simeq 0$ then the effective temperature $T_\text{eff} \simeq 2T$ is twice the thermodynamic surrounding temperature.  We see that in this context, using effective temperatures is a rather drastic multiplicative abbreviation of taking into account the frenetic contribution.\\
The last term in \eqref{kuku} can also be used as indicator of violation of the FDR of the first kind. Or, the difference between the left-hand side and the first term on the right-hand side gives an estimate of the nonequilibrium nature of the reference process.  To make that into a more physical prescription we take the freedom to subtract
\[
\epsilon\left< \,f(x_0)\,\left[ D'_0(\omega) - S'_0(\omega) /2\right]\,\right>_0 = 0
\]
(by normalization) from \eqref{kuku}:  $\langle f(x_t) \rangle_\epsilon - \langle f(x_t) \rangle_0 =$
\[
\epsilon\,\langle f(x_t)\,  S'_0(\omega) \rangle_0 - \epsilon\left< \,[f(x_t)+f(x_0)]\,D'_0(\omega)\right>_0 -\epsilon\,\left<[f(x_t)-f(x_0)]\, S'_0(\omega) /2\,\right>_0
\]
or
\begin{eqnarray}\label{hsa}
\epsilon\left< \,[f(x_t)+f(x_0)]\,D'_0(\omega)\right>_0   &=& -\epsilon\,\left<[f(x_t)-f(x_0)]\, S'_0(\omega) /2\,\right>_0 +\\
&& \{ \epsilon\,\langle f(x_t)\,  S'_0(\omega) \rangle_0 - [\langle f(x_t) \rangle_\epsilon - \langle f(x_0) \rangle_0]\} \label{so}   
\end{eqnarray}
Note that in equilibrium the last line \eqref{so} vanishes because of the Kubo formula \eqref{kub}.  Moreover when $f$ is odd (like a velocity) in the sense that $f(\pi x_0)-f(\pi x_t) = f(x_t)-f(x_0)$ is symmetric under time-reversal, then the right-hand side of the first line \eqref{hsa} also vanishes in equilibrium.  In other words, then, the left-hand side of \eqref{hsa} measures the violation of the Kubo formula (FDR of the first kind).  Now take $f(x) = v$ to get for \eqref{hsa}--\eqref{so}:
\begin{eqnarray}\label{hsa1}
\epsilon\left< \,[v_t+v_0)]\,D'_0(\omega)\right>_0   &=& -\epsilon\,\left<[v_t-v_0]\, S'_0(\omega) /2\,\right>_0 +\\
&& \big\{ \epsilon\,\langle v_t\,  S'_0(\omega) \rangle_0 - [\langle v_t \rangle_\epsilon - \langle v_0 \rangle_0]\big\}    
\end{eqnarray}
In the underdamped regime, see Example \ref{uex}, we can use that the excess entropy flux equals $S'_0= \beta\,\int_0^t\id s\,v_s$ for a constant external perturbation $\epsilon$, so that
$\left<[v_t-v_0]\, S'_0(\omega)\,\right>_0 = \beta\,\int_0^t\id s\left<[v_t-v_0]\,v_s\,\right>_0=0$.  On the other hand, for the excess frenesy we use \eqref{smo},
\[
 D_0' = \frac{\beta}{m\gamma}\int_0^t \id s \,\,F(q_s) - \frac{\beta}{{\gamma}} (v_t - v_0)
 \]
  Hence, for all times $t$,
\begin{eqnarray}\label{has2}
\int_0^t\id s\,\left< \,v_s\, F(q_0)+ v_0\,F(q_s)\,\right>_0  = m\gamma\big\{\int_0^t\id s\,\langle v_s\, v_0 \rangle_0 - \frac{k_BT}{\epsilon} [\langle v_t \rangle_\epsilon - \langle v_0) \rangle_0]\big\}    
\end{eqnarray}
Again, the right-hand side vanishes in equilibrium by the Kubo relation \eqref{kubs}. The left-hand side gives a time-integration of delayed power-dissipation.  For  times $t=\id s$, we see that the frenesy contributes $-F(q_s)\id s + m\id v_s = -m\gamma v_s \id s + \sqrt{2{\cal D}}\,\xi_s$ (multiplied with $\beta/(m\gamma)$) representing the thermostating forces for the unperturbed dynamics.  Together, \eqref{has2} gives a reordering of the linear response around a NESS where the violation of the FDR of the first kind is measured (via the left-hand side) in terms of dissipation. Similar expressions can be obtained by time-modulating the constant $\epsilon \rightarrow \epsilon \cos\nu s$ so that we enter Fourier-space. We can also take the limit $t\uparrow \infty$. The left-hand side then becomes the expectation of the rate of energy dissipation $\langle J\rangle_0$, and we arrive at the Harada--Sasa equality \cite{hasa}, in their notation,
\[
2\pi\,\langle J\rangle = \gamma\int_{-\infty}^\infty[\tilde{C}(\nu) - 2T\,\tilde{R}_S(\nu)]\,\id \nu
\]
The ``tilde'' denotes Fourier-transform and $\tilde{R}_S(\nu)$ is the real part of the transform, $C$ denotes the velocity correlation function and $R$ is the change of velocity caused by a constant external force.\\

After these generalities it is time to get more specific examples.  As for experiments, we refer to \cite{juan} where a driven Brownian particle in a
toroidal optical trap is studied for its linear response of the potential energy.  The frenetic contribution to the response is separately measurable. It shows the experimental feasibility of the entropic--frenetic dichotomy at least for nonequilibrium micron-sized systems with a small number of degrees of freedom
immersed in simple fluids.  For an example with many nonequilibrium degrees of freedom we present a theoretical model as illustration:

\begin{example}[Coupled oscillators]\label{cop}
We put a one-dimensional oscillator $(q_i,p_i)$ at sites $i=1,\ldots,n$ with energy $U=\sum_{i=1}^n \varphi(q_{i+1} - q_i)$ where for example $\varphi(q) =\frac 1{2}q^2 + \frac 1{4}q^4$.  We keep $q_0=q_{n+1}=0$ as boundary conditions.  The dynamics adds white noise $\xi_s(i)$ to every oscillator,
\begin{eqnarray}\label{noisyosc}
\dot{q}_s(i)  &=& p_s(i)\\
\dot{p}_s(i) &=& F_i(q_s) -\frac{\partial U}{\partial q(i)} - \gamma_i p_s(i) + h_s\frac{\partial V}{\partial q(i)} + \sqrt{2{\cal D}}\,\xi_s(i)
\end{eqnarray}
The nonequilibrium resides in the nonconservative forcing $F_i$ and/or in the presence of multiple temperatures $T_i = {\cal D}/(\gamma_i\,k_B)$. A sketch of the situation is depicted in Fig.~\ref{fig:osc}.\\

\begin{figure}[!h]
	\centering
	\includegraphics[width=0.55\textwidth]{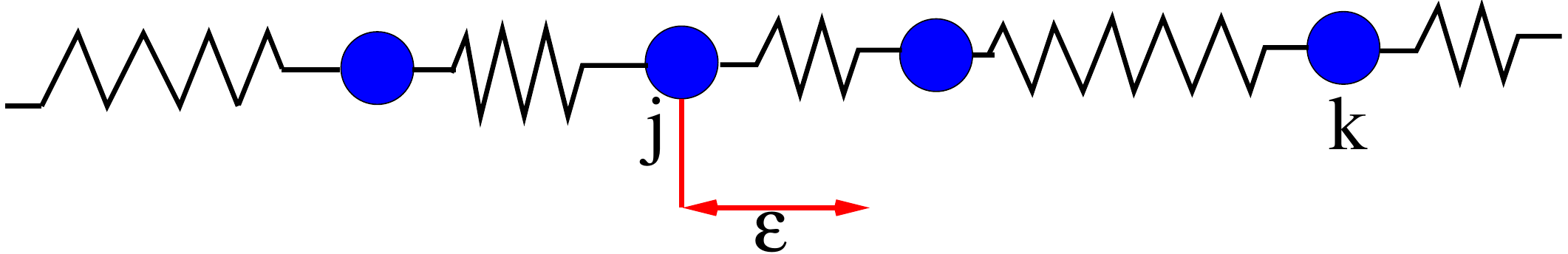}
	\caption{A chain of oscillators may be perturbed by slightly moving a mass at site $j$, applying there a self-potential $V$.  We want to know the effect of the perturbation for the mass at site $k$.  That will be influenced by an existing temperature profile $T_i$.}
	\label{fig:osc}
\end{figure}

 The (small) perturbation is $V(q)$ with amplitude $h_s, s\geq 0$.  Over $[0,t]$ the integrated excess entropy flux is
\[
S = \sum_{i=1}^n \frac 1{T_i}\int_0^t\id s\, h_s \,\frac{\partial V}{\partial q(i)}(q_s)\,p_s(i)
\]
The excess frenesy (in linear order) is
\[
D = \frac{2}{{\cal D}}\sum_i\int_0^t\,h_s\,\frac{\partial V}{\partial q(i)}(q_s)\,\big\{\left[F_i(q_s) - \frac{\partial U}{\partial q(i)}\right]\id s - \id p_s(i)
	\big\}
	\]
As a result (needing some more calculation) we end up with the linear response formula for observable $Q_t$ at time $t$,
\[
\frac{\delta}{\delta h_s}\langle Q_t\rangle_{|_{h=0}} = \sum_i \frac 1{2T_i}\left< \frac{\partial V}{\partial q(i)}(q_s)\,p_s(i)\,Q_t\right>_0 -\langle D\,Q_t\rangle_0
\]
where the last term can be obtained from
\begin{eqnarray}
&&2{\cal D}\langle D\,Q_t\rangle_0 = \sum_i \left< \frac{\partial V}{\partial q(i)}(q_s)\,\left[F(q_s) - \frac{\partial U}{\partial q(i)}(q_s)\right]\,Q_t\right>_0\\
&& -\frac{\id}{\id s}\sum_i \left< \frac{\partial V}{\partial q(i)}(q_s)\,p_s(i) \,Q_t\right>_0 + \sum_{i,j}\left< \frac{\partial^2 V}{\partial q(i)\partial q(j)}(q_s)\,p_j(s)\,p_i(s)\,Q_t\right>_0\nonumber
\end{eqnarray}
As a special case, we take $F=0$, observable $Q = p_k$ and perturbation $V(q) = \epsilon\,q_j$.  We then find the linear response, 
\begin{eqnarray}\label{sus}
\chi_{jk}(t-s) = \frac{\delta}{\delta (\epsilon h_s)}\langle p_t(k)\rangle_{|_{\epsilon=0}} &=& -\left(\frac{\beta_j+\beta_k}{2}\right)\,\langle p_s(j)\,p_t(k)\rangle_0 \nonumber\\
-&& \frac{1}{2{\cal D}}\left(\left<\frac{\partial U}{\partial q(j)}(q_s)\,p_t(k)\right>_0 + \left<p_s(j)\,\frac{\partial U}{\partial q(k)}(q_t)\right>_0\right)
\end{eqnarray}

\begin{figure}[!h]
		\centering
	\includegraphics[width=0.65\textwidth]{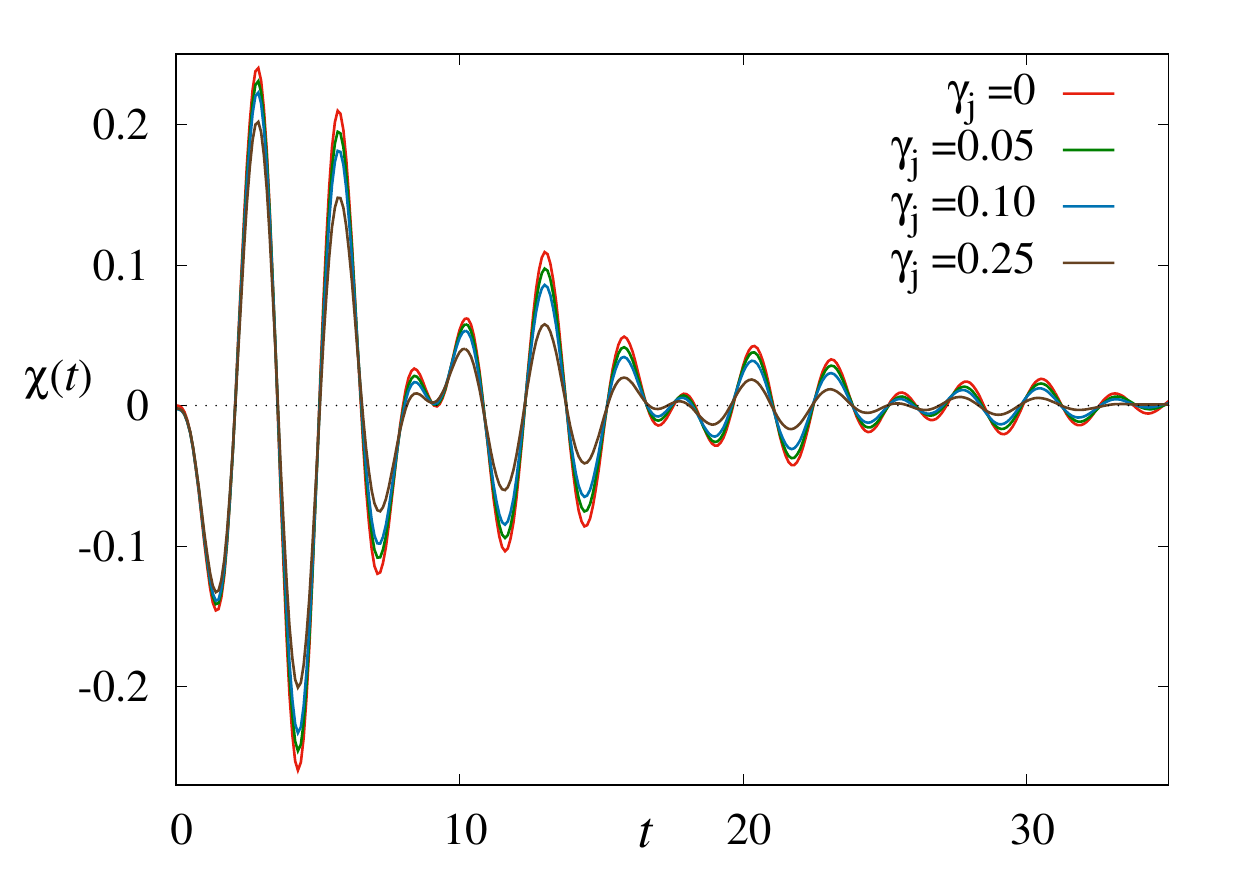}
\caption{The susceptibility \eqref{sus} of $p(k)$ for perturbing at $q(j)$, for different values of $\gamma_j$ with $T_j=1.5$ and $\gamma_1=\gamma_n=1,$ $T_1=2$ and $T_n=1$; all other $\gamma_i\equiv = 0$.  A small perturbation of the mass at site $j$ causes a damped oscillatory movement of the mass at site $k$.  Interestingly, the limit in which the damping $\gamma_j \downarrow 0$ makes sense, erasing the thermal noise in the bulk. The plot refers to the dynamics \eqref{noisyosc} with force $F_i= -2q(i)+q(i-1)+q(i+1) - \alpha q(i)^2 -k q(i)$ with $\alpha=1.0, k=2.0$ and $U=0$.  Figure courtesy of Urna Basu.}
	\label{fig:g1}
\end{figure}

Observe the spacetime reciprocity $j\leftrightarrow k, s\leftrightarrow t$. In Fig.~\ref{fig:g1} we see the susceptibility $\chi_{jk}(t-s)$ as function of time for different values of the damping $\gamma_j$.  It appears that the limit of vanishing bulk thermal noise continues to make sense for the response, \cite{priv}.  That example thus stands for the study of longitudinal waves in heat conducting strings.   
\end{example}

\begin{example}[Linear response of jump processes]
	We revisit the Markov jump processes of Section \ref{mjp}, with the parametrization \eqref{gfo}; see also \cite{mp}.
	We take a perturbation 
	\begin{equation}\label{pra}
	s(x,y)\rightarrow s(x,y) + \epsilon\,s_1(x,y),\quad  a(x,y)\rightarrow a(x,y) + \epsilon\,a_1(x,y)%\nonumber\\
	%\gamma(x,y) =&& 1+ \epsilon\varphi(x,y),\quad F(x,y) = \epsilon a(x,y)
	\end{equation}
	to linear order in $\epsilon$. Then, the excess frenesy equals
	\begin{equation}\label{edal}
	D(\omega) = -\epsilon\sum_s a_1(x_{s^-},x_s) + \epsilon\int_0^t \id s \sum_y k(x_s,y)[a_1(x_s,y) + \frac 1{2} s_1(x_s,y)]
	\end{equation} 
	and 
	\begin{eqnarray}\label{linre}
	\langle O\rangle_\epsilon - \langle O\rangle_0 &=& \frac{\epsilon}{2}\, \left < \sum_s s_1(x_{s^-},x_s)\,O(\omega)\right >_0\\
	&+& \epsilon\;\left <\left[\sum_s a_1(x_{s^-},x_s) - \int_0^t \id s \sum_y k(x_s,y)[a_1(x_s,y) + \frac 1{2} s_1(x_s,y)]\right]\,O(\omega)\right >_0 \nonumber
	\end{eqnarray}
	gives the response for an arbitrary path-observable $O$ over time $[0,t]$ in terms of a reference nonequilibrium condition.  The first term on the right-hand side of \eqref{linre} is proportional to the correlation of the entropy flux $S$ with the observable $O$ and gives rise to the usual Kubo-formula \eqref{kubo} with the time-derivative when the perturbation is caused by a potential; see \cite{mp}.\\
	
	Example \ref{rw} is the simplest illustration of the above\footnote{With the possible abuse of notation that there $\epsilon$ stands for the nonequilibrium driving, and we perturb $\epsilon\rightarrow \epsilon + \id \epsilon$.}, where we perturb around a fixed (large) value of $\epsilon$.  The current appears in \eqref{ve} and its derivative equals
	\begin{equation}\label{vel}
	\frac 1{L}\frac{\id}{\id \epsilon} \langle v\rangle_F =  2a'(\epsilon)\,\sinh \frac{\epsilon}{2} + a(\epsilon)\,\cosh\frac{\epsilon}{2} \simeq [a'(\epsilon)+ a(\epsilon)/2]
	\,e^{\epsilon/2}
	\end{equation}
	The derivative $a'(\epsilon)$ only contributes for  $\epsilon \neq 0$.  The negativity of $a'/a (\epsilon) < -1/2$ for large $\epsilon$ implies a negative differential conductivity. The same can be concluded from taking the derivative of \eqref{redec}, which is reproducing \eqref{vel} with $\langle J\,;\,J\rangle_0 \simeq \langle N\,;\,J\rangle_0 \simeq t\,a(\epsilon)\exp\epsilon/2$.
	\end{example}
	
Such a simple scenario as above with the crucial role of the frenetic contribution gets realized in more examples, including responses to temperature and chemical affinities; see \cite{sar,zia,gar,negheatcap,oliver,chemfalasco,hao}.  To pick one, in \cite{hao} one sees modifier activation--inhibition switching in enzyme kinetics. A more abstract scenario (going beyond the case of Markov jump processes) goes as follows:  taking the observable $O=S'_0$ (typically proportional to a current), linear order response gives
	\[
	\langle S'_0\rangle_\epsilon - \langle S'_0\rangle_0 = \frac{\epsilon}{2} \langle (S'_0)^2\rangle_0 - \epsilon \langle S'_0\,D'_0\rangle_0 
	\]
In contrast with \eqref{sf}, a positive correlation between the linear excesses in entropy flux and in frenesy in the {\it original} dynamics yields a negative frenetic contribution.  In and close-to-equilibrium, $\langle S'_0\rangle_\epsilon - \langle S'_0\rangle_0\geq 0$ always.   Two necessary conditions for a negative susceptibility for the observable $S'_0$ are, (1) one needs to be sufficiently away from equilibrium, and (2) one needs a positive correlation 
	$\langle S'_0\,D'_0\rangle_0 >0$ in the original process.  More generally, it is the frenetic contribution that can make currents to saturate and provide homeostatic effects far enough from equilibrium.\\

	We also recall an application of the Cram\'er--Rao bound, which enables to give a general bound on response functions.  That was exploited in the 
	Dechant-Sasa inequality \cite{des} to give that
	\[
	\left(\frac{\partial \langle O \rangle_\epsilon
	}{\partial\epsilon}\big|_{\epsilon=0}\right)^2 \leq 2\,\text{Var}[O]\,\langle A''_0\rangle
	\]
	for an arbitrary path-observable $O=O(\omega)$ on $[0,t]$ with variance Var$[O]$; see \cite{des,terl} for details. Naturally, the (unperturbed) expectation $\langle A''_0\rangle$ is related to the frenesy.\\

As a final remark, nonequilibrium linear response as formalized above can also be used for an expansion of the stationary distribution around a reference nonequilibrium.  In particular we mention the work of Komatsu and Nakagawa in \cite{naoko} for characterizing nonequilibrium stationary distributions.  A similar analysis followed in \cite{col,nongrad}.  Work remains to be done towards applications on population selection and the understanding of relations with interdisciplinary aspects having to do with trophic levels in foodwebs or with the appearance of homeostasis in biological conditions, to mention just two.

\subsubsection{Modified (Sutherland--)Einstein relations}
\begin{figure}[!h]
	\centering
	\includegraphics[width=0.45\textwidth]{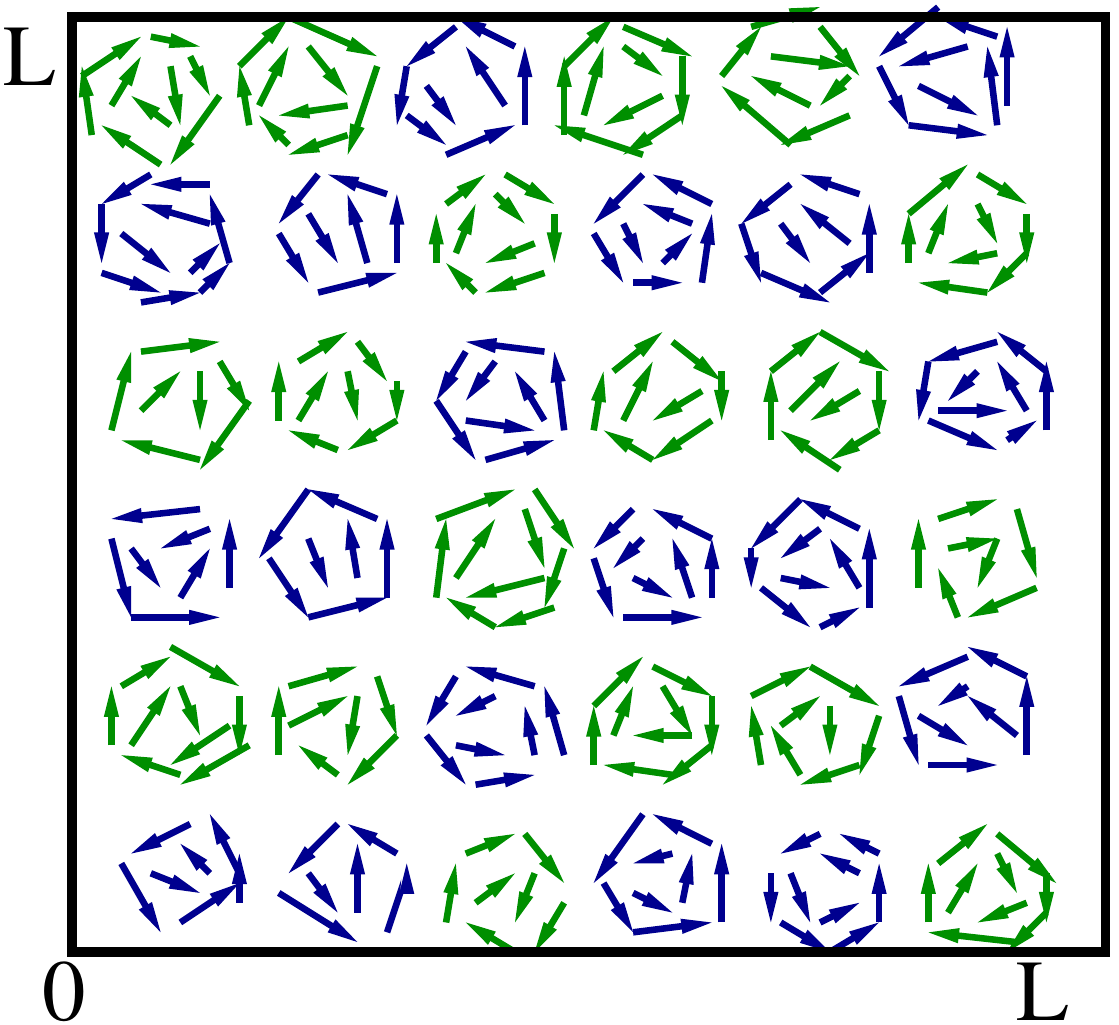}
	\caption{A nonconservative periodic force field for which the Sutherland-Einstein relation gets modified.}
	\label{fig:rot}
\end{figure}

Around nonequilibrium, the FDR of the first kind (between mobility and diffusion) is violated, and the Sutherland--Einstein relation must be corrected with a frenetic contribution.  We refer to the constructions in \cite{proc,soghra,gal} for more introduction and examples.\\
In general, we take a particle of mass $m$ in a heat bath according to the Langevin dynamics for the position $\vec{r}_t$  and the velocity $\vec{v}_t$,
\begin{eqnarray}
\dot{\vec{r}}_t  &=& \vec{v}_t \label{general}\\
m\dot{\vec{v}}_t &=& \vec{F}(\vec{r}_t)-\gamma m \vec{v}_t + \sqrt{2m\gamma\,k_B T}\,\vec{\xi}_t \nonumber
\end{eqnarray}
We get out of equilibrium when the force $\vec F$ is not derived from a periodic potential. It can be arbitrarily large. We have no confining potential and no global bias, meaning that the steady (net) velocity is zero. The easiest is to work with a spatially periodic force field $\vec{F}$ which adds vortices in its rotational component, e.g. a lattice of convective cells as in Fig.~\ref{fig:rot}.  

\noindent The vector $\vec{\xi}_t$ is standard Gaussian white noise.\\
When the system is not in equilibrium, and we search for an expression for the mobility \eqref{mobi}, we can use  \eqref{linwon} or \eqref{kuku} where the perturbation changes $\vec{F}(\vec{r}_t)\rightarrow \vec{F}(\vec{r}_t) + {\cal E}$.  We look at the linear response in ${\cal E}$.  Frenetic terms show up so that the mobility and diffusion constants \eqref{diffi} are no longer proportional.  See \cite{proc} for a detailed derivation of the following result:  the nonequilibrium modification of the Sutherland-Einstein relation is given by
\begin{eqnarray}
&& M_{ij} = \frac{1}{k_BT}{\cal D}_{ij} - \lim_{t\to\infty}\frac{1}{2\gamma m\,k_BT \,t}\,\int_0^t \id s\,\Big<\frac{({\vec r}_t - {\vec r}_0)_i}{t};F_j(\vec{r}_s,\vec{v}_s)\Big>_0\label{genresult5}
\end{eqnarray}
(notation from \eqref{mobi}--\eqref{diffi}.)  The frenetic contribution gives a spacetime correlation between applied forcing and displacement (last expectation in the right-hand side of \eqref{genresult5}).  Quite generally, the diffusion is much more sensitive to the strength of the force than is the mobility. The deviation with respect to the Sutherland-Einstein relation is second order in the nonequilibrium driving.  We refer also \cite{sar3} for further analysis and phenomenoloby, including the occurrence of negative mobilities.\\

The formula \eqref{genresult5} is again similar to a Harada-Sasa equality (see \eqref{has2} and formula 22 in \cite{hasa}).  It also invites some inverse problem. In the paper \cite{gal} the theory of linear response around nonequilibria is used to probe active forces in
living cells: by measuring the force, one obtains the correlation between force and displacement which is exactly the frenetic part in \eqref{genresult5}.\\

\begin{figure}[!h]
	\centering
	\includegraphics[width=0.45\textwidth]{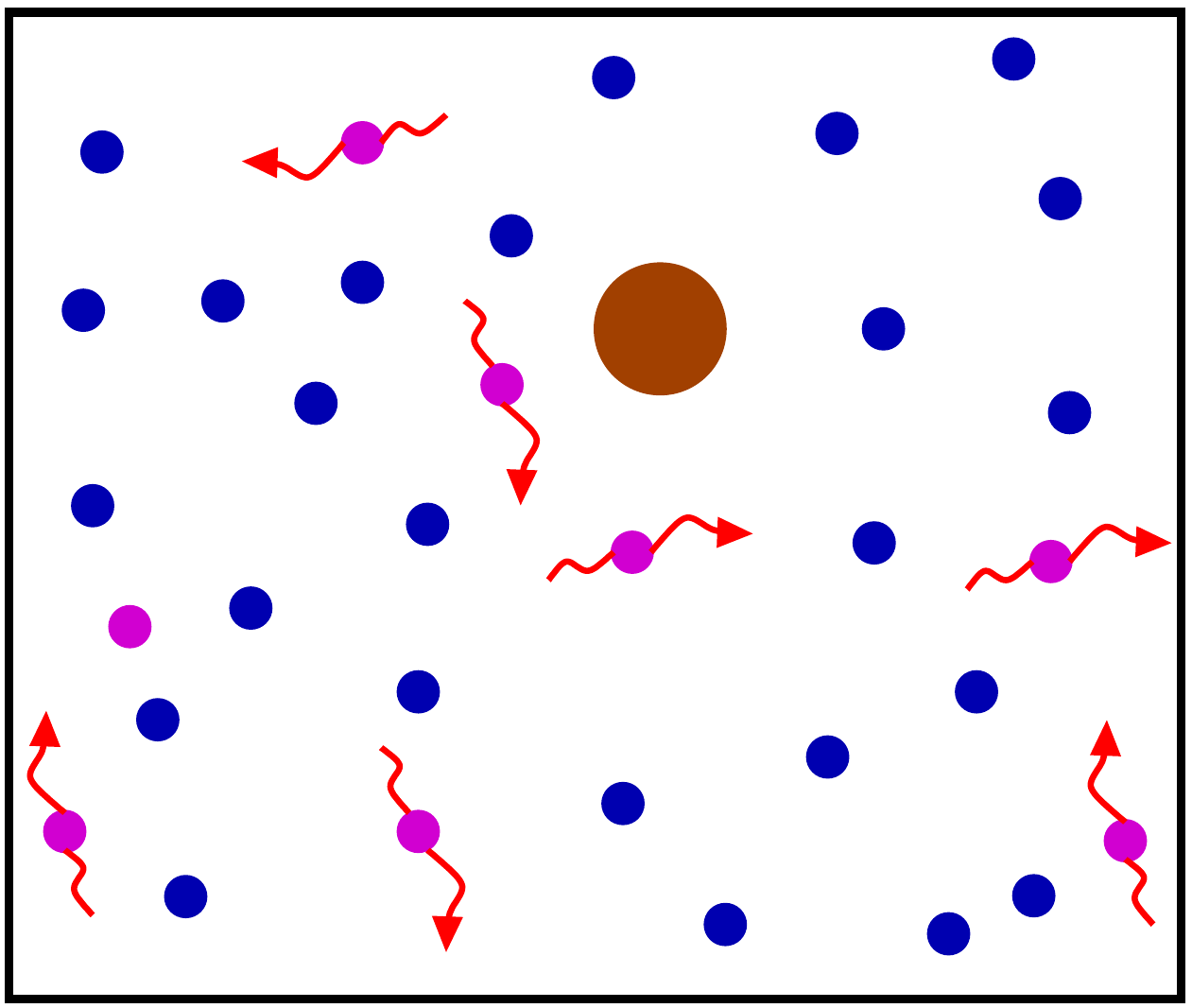}
	\caption{Cartoon of three levels: slow probe, faster nonequilibrium medium and and even faster thermal bath. The probe motion perturbs its environment. The response of the medium is needed to derive the fluctuating dynamics of the probe.}
	\label{fig:swi}
\end{figure}

To understand the modifications to the Einstein relation (FDR of the second kind) we must revisit the calculations in Section \ref{kind}. The set up remains the same; see Fig~\ref{fig:swi}. The logic remains the same as well but we must add the frenetic contribution to \eqref{fob}.  It means that the induced friction gets a modification (and is no longer purely dissipative into the environment) because of the nonequilibrium nature of the bath.  For details we refer to \cite{jsp,stefan,leipzig,krueger}, where \cite{leipzig} also discussed the possible changes in the noise statistics related to the nonequilibrium bath.

\subsubsection{Active particles: NO local detailed balance}\label{nl}
To show how the formalities proceed even in the absence of local detailed balance, we give here the example of linear response for an active particle system. See for example \cite{mar} for a general review on active particles.\\

 We start by illustrating the situation in the case of an active Ornstein--Uhlenbeck (AOU) particle , \cite{10}.
 Linear response for AOU particles has been subject of various papers already, including \cite{cip,pao}.\\
Consider a particle in one dimension in a potential $V$ and with position $q_s$ following
\begin{equation}\label{aousp}
\dot q_s = \cal E\,v_s  - \mu V'(q_s) + \mu\,h_s;\qquad \tau\dot v_s + v_s = \sqrt{2R}\,\xi_s
\end{equation}
The noise is $v_s$ and while it is mean-zero Gaussian, it is not white.  In fact, 
\begin{equation}\label{ga}
\gamma(s-s') := \langle v_s\,v_{s'}\rangle = \frac{R}{\tau}\,e^{-|s-s'|/\tau}  \stackrel{\tau\downarrow 0}{\longrightarrow} 2R\,\delta(s-s')
\end{equation}
The time-constant $\tau$ measures the persistence time in the process $v_s$, which is then applied as an external field with amplitude $\cal E$ to the particle motion.  As a consequence, the process $(q_s)$ is not Markovian and is not satisfying the FDR of the second kind (Einstein relation), in contrast with all the examples in Section \ref{decoex}. 

 For $\tau=0$ the motion is passive with standard white noise $\xi_s$ of strength $R$.  The Einstein relation would then set $R{\cal E}^2= k_BT\,\mu$ where $\mu$ is the mobility.  We have already added a constant perturbation to \eqref{aousp}, with time-dependent amplitude $h_s$.  The question is again to understand the linear response 
\[
\langle O\rangle_h - \langle O\rangle_0
\]
for some observable $O$ in the perturbed ensemble $\langle\cdot\rangle_h$ with respect to the original (unperturbed) $\langle \cdot\rangle_0$.\\
Even though the model does not satisfy local detailed balance (of Section \ref{lodetb}), we can still apply the same response formul{\ae} if we identify the action in \eqref{aha} to apply \eqref{peens}.  In a formal sense, the probability of a trajectory $\omega$ of positions $q_s, s\in [-\infty,+\infty]$, is proportional to
	\begin{equation}\label{ker}
	\text{Prob}_h[\omega] \propto \exp - \frac 1{2} \int \id s \int\id s'\,\Gamma(s-s') v_s\, v_{s'}
	\end{equation}
if we substitute
	\[
	v_s = \frac 1{\cal E}\left( \dot q_s + \mu V'(q_s) - \mu\,h_s  \right)
	\]
and use the symmetric kernel $\Gamma(s)$ for which
\[
\int \id s'\,\Gamma(s-s') \gamma(s'-s'') = \delta(s-s'')
\]
Via Fourier transform\footnote{We can also verify directly by using $\partial_{xx}^2 e^{-\alpha |x|} = -2 \alpha\, \delta(x) \,e^{-\alpha |x|} + \alpha^2 \,e^{-\alpha|x|}$.} it is straightforward to get $\Gamma(s) = [\delta(s) - \tau^2\, \ddot\delta(s)]/(2R)$.\\
As usual we put
\[
\text{Prob}_h[\omega] = e^{-\cal A}\,\text{Prob}_0[\omega]
\] and find the action
\begin{eqnarray}\label{ah}
{\cal A} &=& -\frac 1{\cal E^2}\int \id s\, h_s\int\id s'\, \Gamma(s-s')\,\big( \dot q_{s'} + \mu V'(q_{s'}) \big) + O(h^2)\nonumber\\
&=&  -\frac 1{2\cal E^2}\int\id s'\, K(s')\,\left( \dot q_{s'} + \mu V'(q_{s'}))\right) + O(h^2)
\end{eqnarray}
where the kernel $K_s := h_s - \tau^2\ddot{h}_s$.\\
Concerning the nature of the stochastic integral \eqref{ah} it is interesting to remark that there is no difference here between the It\^o and the Stratonovich convention. For the first term in the integral of \eqref{ah} we can write
\[
I:=\int\id s'\, K(s')\,\dot q_{s'} \simeq \sum_{s'} K(s')\, (q(s'+\delta) - q(s'))
\]
where the integral is discretized to become a sum where the difference between consecutive $s'$ is of order $\delta$.  For the time-symmetric part $I\theta + I$ (and also time-reversing the perturbation), we see that
\begin{eqnarray}\label{alp}
I + I\theta &\simeq& \sum_{s'} \left(K(s')\, (q(s'+\delta) - q(s'))+ K(-s')\, (q(-s'-\delta) -q(-s'))\right)\nonumber\\
&=& \sum_{s'} \left(K(s')\, (q(s'+\delta) - q(s'))+ K(s')\, (q(s'-\delta) - q(s'))\right)\nonumber\\
&=&\sum_{s'} [K(s') - K(s'+\delta)]\,(q(s'+\delta)-q(s'))
\end{eqnarray}
which tends to zero as $\delta\downarrow 0$.  There is indeed no short-time diffusion and the behavior of $q_s$ is ballistic for every $\tau>0$.  The excess frenesy as induced by the perturbation to linear order, is therefore
\begin{equation}\label{d}\frac1{2}\big({\cal A}\theta + \cal A\big) = D =   -\frac{\mu}{2\cal E^2}\int\id s\, K(s)\, V'(q_s) \end{equation}
On the other hand, the time-antisymmetric part of the action is
\begin{equation}\label{s}
{\cal A}\theta - \cal A = \frac 1{\cal E^2}\int\, K(s)\, \dot q_s\,\id s
\end{equation}
In the passive case where $K(s) =\mu h_{s}/R$ local detailed balance would impose $\mu \cal E^2 R = T$ to be the temperature and \eqref{s} would represent the entropy flux per $k_B$. In the active case, we can only consider $R\,\cal E^2$ as a measure of the strength of dynamical activity delivered by the Ornstein-Uhlenbeck noise.  There is however no physical identification of ${\cal A}\theta - \cal A$ with the (excess) entropy flux due to the perturbation.\\ 
Nevertheless, the formula of response to linear order holds unchanged as
\begin{equation}\label{plinr}
\langle f(q_t)\rangle_h - \langle f(q_t)\rangle_0= \frac 1{2\cal E^2}\int^t\id s\, K_s\left< f(q_t)\,;\,\big( \dot q_s + \mu V'(q_s)\big) \right>_0
\end{equation}
for functions $f$ and with $K_s = h_s -\tau^2\ddot{h}_s$.  That second term, proportional to the persistence time, induces a double time-derivative to apply on the expectation, of course also depending on $\tau$. \\

A second example of an active particle model is the well-known run-and-tumble process, also called Kac or telegraph process \cite{wei}, where the particle moves on the real line with positions $q_s$ following
\begin{equation}\label{nzerot}
\dot{q}_s = c\,\sigma_s + \sqrt{2T} \xi_s,\; \,\quad  \sigma_s \longrightarrow -\sigma_s \text{ at rate }  a
\end{equation}
where the noise $\sigma_s= \pm 1$ is dichotomous.  Again, there is no local detailed balance, and no presence of an Einstein equation except in the limit $a\uparrow \infty$ where the noise becomes statistically indistinguishable from being white.
That is a finite temperature ($T$-)generalization of the usual run-and-tumble process introduced in \cite{act1}; see also \cite{cv2}.  The Smoluchowski equation for the spatial density $\rho = \rho(q,t)$ satisfies
\begin{equation} \label{T-tel}
(\partial_t -T\partial_q^2)^2\rho - c^2\partial_q^2\rho = -2a(\partial_t-T\partial_q^2 )\rho
\end{equation}
The derivation of \eqref{T-tel}, a thermal telegraph equation, is done in \cite{act1}.\\
We start the process  at $q=0$ with equal probability of having $\sigma_0 =1$ or $\sigma_0 =-1$.  We find $\langle q^2_t\rangle_0$ for large $t$  by multiplying equation \eqref{T-tel} by $q^2$ and integrating:
\[
\ddot{\langle q^2\rangle_0}-2c^2=-2a\dot{\langle q^2\rangle_0}+4aT
\]
Therefore, the diffusion constant is
\begin{equation}\label{diffu}
{\cal D}:=\lim_{t \to \infty}\frac{\langle q^2_t\rangle_0}{2t}
= T + \frac{c^2}{2a}
\end{equation}
(see also \cite{ind}).  Note that there is already diffusion at zero temperature $T=0$.\\
To get the mean velocity  $v = \lim_{t \to \infty} \langle q_t\rangle_\epsilon/t$ resulting from the application of an extra external field $\epsilon $, we modify in \eqref{nzerot} the drift $\sigma_s\,c\rightarrow \sigma_s\,c + \epsilon$.  We easily find that  $v = \epsilon$ and the mobility is thus ${\cal M}=1$.
Per consequence,
\begin{equation} \label{Suth}
\frac{\cal D}{T} = 1 +\frac{c^2}{2a \,T} > \cal M
% \frac{{\cal D}}{\mu T_\text{eff}} = \frac{T+\frac{c^2}{2a}}{T+\frac{c^2}{\kappa+2a}}> 
\end{equation}
and the Sutherland-Einstein relation is broken.  See more discussion in \cite{act1}.
The Sutherland--Einstein relation has been discussed as well for active systems with a possible interpretation  in terms of an effective temperature in \cite{Berthier,Szamel}.

\subsubsection{Open problems}\label{open}
We mention a couple of natural open problems related to response around nonequilibria.

\begin{enumerate}
	\item 
\underline{Singular response}:  In the basic formula \eqref{aha} for relating two dynamical ensembles, we assume implicitly that the set of allowed trajectories are the same for both; only the weights change. In mathematial terms, we speak of mutual \emph{absolute continuity} of the processes, as part of the hypothesis in the Girsanov theorem \cite{girs}.  For various classes of dynamics that assumption is not satisfied at first sight.  There may be various reasons, and we very briefly discuss three.  When we consider two Markov diffusions with different noise strengths, then they are not comparable.  That happens in particular for changes in temperature.  So, at first sight there is a fundamental problem with thermal response, how a change in temperature changes the expectations.  That question has been treated from various sides, for different questions and with different methods.  We refer to \cite{t1,t2,t3,t4,t5,t6} for some of the progress.  A second case of possible problems arises when trajectories are subject to deterministic constraints, which are perturbed.  Again, trajectories become incomparable. For instance, in the Example \ref{cop} we have added noise to each oscillator.  Perturbing the chain in a region without noise, where the dynamics is purely Hamiltonian creates problems for the method with dynamical ensembles.  Of course, for the linear response around equilibrium, there is no problem because we know the (stationary) equilibrium distribution and there the Agarwal formula \cite{ag} (see also \cite{njp}) can be used.  In the same paper \cite{njp} and via the same method a linear response for dynamical systems is illustrated.  A third (always) related case is that of changes in geometry and topology.  Nonequilibrium may be a topological effect as e.g. allowing circuits is essential for breaking detailed balance.  Again, changes in the network architecture or topology may be give rise to incomparable trajectories.\\
In general, stochastic regularization is a good method to pragmatically deal with it, if linear response makes sense at all.  That is illustrated in Example \ref{cop} and in Fig.~\ref{fig:osc} for a chain of oscillators where the dynamics becomes Hamiltonian in the bulk.
\item
 
\underline{Many-body physics}:  We have emphasized since the start that response expansions must be useful.  That means also that the observables appearing in the expectations of linear or nonlinear response should be measurable.  Today, much progress was made to follow trajectories of individual particles.  The many-body case is however still very challenging.  There seems no good escape here; frenesy is necessary in response around nonequilibria and involves many-body kinetics.  Other relations avoid the details of response but still give useful relations.  We have in mind for example the discussed Harada-Sasa equality where the energy dissipation is obtained from experimentally accessible quantities alone, without knowing every detail of the system.  Again, physical coarse-graining towards more reduced descriptions appears a good option; see e.g. \cite{urna18}.
 
 \item
 \underline{No local detailed balance}:  We have supposed throughout that we work under the condition of local detailed balance. That is not a strict mathematical prerequisite, but it is essential for the physical interpretations.  In Section \ref{nl} we have seen the examples of linear response for active Ornstein-Uhlenbeck and run-and-tumble processes.  Those were the easy cases however.  Extensions of the FDR of the first and the second kind for active systems which are in direct contact with nonequilibrium degrees of freedom are therefore to be explored further.  We have seen how the Einstein relation between noise and friction gets modified for probes coupled to nonequilibrium reservoirs, but much needs to be clarified here for benchmarking a physically motivated active Brownian motion.
 Active systems as we encounter in biological processes break the FDR, and we wish to construct the response from the tools of the present paper.  See e.g. \cite{rol} for such a challenge.
 
 \item 
\underline{Quantum nonequilibrium}:  The linear response around quantum nonequilibria faces various problems,  To start, we lack good modeling of quantum nonequilibrium processes\footnote{Obviously, we do have a number of powerful computational models in quantum nonequilibrium physics, as provided from the Schwinger--Keldysh formalism \cite{sw,ke} or from Feynman-Vernon theory \cite{fey}. We do not include them here in the discussion, as our ambition is to attempt a trajectory-based approach.}.  Quantum open dynamics is usually treated in the weak coupling limit where Markov approximations arise.  It is however not so clear whether true quantum phenomena (e.g. outside the Coulomb blockade regime for quantum dots) can be modeled physically correctly by Markov dynamics.  Entanglement between system and reservoir or between reservoirs is probably necessary.  Dynamics such as via Lindblad evolutions have fast decoherence in the energy basis and can only be approximately touching the quantum world.  A second problem has to do with a quantum notion of dynamical activity.  A trajectory-based approach for open quantum systems does not appear straightforward.  We have little idea for example whether a small particle subject to zero-point quantum fluctuations (only) will undergo a diffusive (or very subdiffusive) motion, see \cite{sorkin} for an exciting possibility based on the quantum FDR.  We continue this discussion in Section \ref{qca}.

\item
\underline{Ageing and glassy systems}:
This review does not deal explicitly with response theory in disordered and glassy systems,\cite{hen}.  That is unfortunate as one of the main forces for the development of response theory out-of-equilibrium has indeed been the physics of glassy systems; see e.g.~\cite{ric,bou}. The focus of this review is much more on response around steady behavior, while glasses refer to a transient albeit very long-lived condition.  The methods of the previous Sections remain valid but the nonequilibrium sits entirely in the nature of the condition with a dynamics that is, for the rest, undriven and satisfying detailed balance with respect to an asymptotic equilibrium. While the physics is clearly much more complicated than what has been presented in the majority of examples so far, there is a further good reason why it should appear as an (advanced) application of response theory in a trajectory-based approach.  Today, there is a growing trend to emphasize the kinetics of glassy behavior, instead of the thermodynamics of metastability.  The general idea is that many-body interactions create kinetic constraints for the evolution and relaxation to equilibrium.  But that is exactly in line with the frenetic aspects we have been enphasizing: relaxation requires the possibility of traffic between mesoscopic conditions.  We have seen examples of particle transport where the current gets strongly diminished when pushing harder as the frenesy takes over as the main component in the action.  Similarly, people have considered the glassy phases and transitions as manifestations of jamming and transitions in dynamical activity \cite{gar,lip}.

\item
\underline{Applications and experiments}:
While we tried to emphasize the importance of the frenetic contribution to response, there are clearly many more applications and insights that can be gained; see also \cite{springer}.  One possible avenue is to understand better what determines the scale of susceptibilities.  How sensing works, in other words.
It would for example be interesting to understand the validity of the Weber-Fechner law (1834) from psychophysics and which states that the relationship between stimulus and perception is logarithmic; see e.g. \cite{port}.\\
We see also that weak susceptibility of certain observables (homeostasis) would follow from near orthogonality of the observable $O$ and the excess action,
$O \perp  \left[-D'_0 + \frac 1{2} S'_0 \right]$, in the sense of a vanishing right-hand side in \eqref{linwon}.  Such points of zero susceptibility are reached when moving from a regime of positive to negative susceptibility.\\
At the same time, experiments on measuring the role of frenesy are still limited.  Trajectory-based response is feasible with the newest tools of tracking and data selection.  We hope more of that can be used for understanding nonequilibrium response.

\end{enumerate}   

\subsection{Nonlinear response around equilibrium}\label{2or}

One may wonder whether the (mutilated) ensemble \eqref{mut} or just the fluctuation identity \eqref{locd} would suffice to continue response theory to second order.
It was explained in \cite{maarten} why that does not work.  If all we know is \eqref{locd} (the basis for all fluctuation theorems), then, equivalently, in the nonequilibrium process,
\[
\langle O\rangle_\epsilon = \langle O\theta\,e ^{S}\rangle_\epsilon\]
Apply that to a time-symmetric observable, $O=O\theta$ and expand to linear order in the nonequilibium strength $\epsilon$:
\[
\langle O\rangle_\epsilon = \langle O\,e ^{S}\rangle_\epsilon\implies \langle O\rangle_\epsilon = \langle O\rangle_\epsilon- \epsilon\langle OS'_0\rangle_\text{eq} =
\langle O\rangle_\epsilon\]
which is empty.  Linear response around equilibrium follows from fluctuation theorems (i.e., identities like \eqref{locd}) \emph{only} for time-antisymmetric observable, like for showing Green-Kubo relations.  It implies that second order response, even for antisymmetric obervables, does not fly.  We need another method.\\

The question of nonlinear response around equilibrium has of course been considered in many important papers.  We mention 
\cite{bou} for the context of disordered systems to enable measurement of a correlation length and \cite{lip} where the frenetic term plays a central role.\\

Section \ref{are} can be continued from \eqref{wonders}.  We start again with the equilibrium reference with expectations $\langle\cdot\rangle_\text{eq}$.  We suppose that $ S = \epsilon \,S'_0$, meaning that the entropy flux determines the order of the perturbation, e.g. from adding external fields or potentials as perturbations.  Using \eqref{wonders} with $S_0''=0$ and since both $D_0''$ and $(S_0')^2$ are symmetric under time-reversal,
\begin{equation}\label{2n4}
\langle O - O\theta\rangle_\ep = \epsilon \,\langle S'_0(\omega)\, O(\omega)
\rangle_{\text{eq}} - \epsilon^2\,\langle D'_0(\omega)\,S'_0(\omega)\, O(\omega)\rangle_{\text{eq}}
\end{equation}
With a state function $O(\omega) = f(x_t)$, applying formula \eqref{2n4}, we get the next order beyond the traditional Kubo formula \eqref{kubs},
\begin{equation}\label{kubo2}
\langle f(x_t)\rangle_\ep - \langle f(x_t)\rangle_{\text{eq}} =  \varepsilon\,\langle S'_0(\omega)\,f(x_t)
\rangle_{\text{eq}} - \varepsilon^2\,\langle D'_0(\omega) \,S'_0(\omega)\, f(x_t)\rangle_{\text{eq}}
\end{equation}
We have used again that  $\langle f(\pi x_0)\rangle_{\text{eq}} = \langle f(x_0)\rangle_{\text{eq}} = \langle f(x_t)\rangle_{\text{eq}}$.  The result \eqref{kubo2} is valid for general time-dependent
perturbation protocols as well; see \cite{pccp}.\\
To extend the Green--Kubo formula \eqref{gka}, we take an antisymmetric observable $O(\theta\omega) = -O(\omega)$ as  for time-integrated particle or energy currents, $O(\omega) = J(\omega)$.  Then, from \eqref{2n4},
\begin{equation}\label{gk}
\langle J\rangle_\ep = \frac{\varepsilon}{2} \,\langle S'_0(\omega)\, J(\omega)
\rangle_{\text{eq}} - \frac{\varepsilon^2}{2}\,\left < D'_0(\omega)\,S'_0(\omega)\, J(\omega)\right >_{\text{eq}}
\end{equation}
Similarly, taking $O(\omega) = S'_0(\omega)$ in \eqref{2n4} makes
\[
\langle S'_0 \rangle_\ep = \frac{\varepsilon}{2} \,\left < (S'_0)^2
\right >_{\text{eq}} - \frac{\varepsilon^2}{2}\,\left < D'_0\,(S'_0)^2\right >_{\text{eq}}
\]
so that the sign of the second-order term depends on an entropy--frenesy correlation in equilibrium, correcting the FDR \eqref{sf}.\\

Starting the discussion of the next section it is interesting to observe that perturbations which are thermodynamically equivalent (having the same $S'_0$), still yield a different response.  That is due to the frenetic contribution (different $D'_0$).
Sensing beyond close-to-equilibrium is a kinetic effect; see Fig.~\ref{fig:sad}. 

\subsubsection{Feeling kinetics}\label{resk}

Suppose we have a gas in a volume $V$ which is open to exchange of particles from a chemical bath at temperature $T$ and chemical potential $\mu$. The gas finds itself in thermal and chemical equilibrium with fixed volume, chemical potential and temperature. Of course, the number  $N(t)$ of particles at time $t$ is variable. The density $\langle N\rangle_\text{eq}/V$ is constant and determined by the environment $(\mu,T)$. That is the preparation at time zero.  Let us then change the chemical potential from $\mu$ to $\mu + \delta$ at fixed $T$, for some small $\delta$.  In time the gas will relax to the new equilibrium at $(\mu+\delta,T)$, with an evolution of the density through the expected particle number $\langle N(t)\rangle$.  Its change in time is given by response theory.  In the linear regime, from \eqref{kub}, we get
\[
\langle N(t) \rangle - \langle N\rangle_\text{eq} = \beta\delta\,\int_0^t \id s\,\big< J(s);N(t)\big>_\text{eq} 
\]
Here, $\langle\,\cdot\,\rangle_{\text{eq}}$ is the expectation in the original equilibrium process with $(\mu,T)$, and $J(s)$ is the net current at time $s$ of particles entering the environment. Using $\int_0^t\id s J(s) = N(t) - N(0)$, we see
\[
\langle N(t) \rangle - \langle N\rangle_\text{eq} = \frac{\beta\delta}{2}\,\left< [N(t)-N(0)]^2\right>_\text{eq} 
\]
which is the FDR of the first kind (linear in small $\delta$).  We only used \eqref{kub} and a general thermodynamic description in terms of particle number, entropy flux and the relevant intensive variables.  The  expectation takes care of the rest. The expectation in the right-hand side only depends on the original chemical potential.

\begin{figure}[!h]
	\centering
	\includegraphics[width=0.65\textwidth]{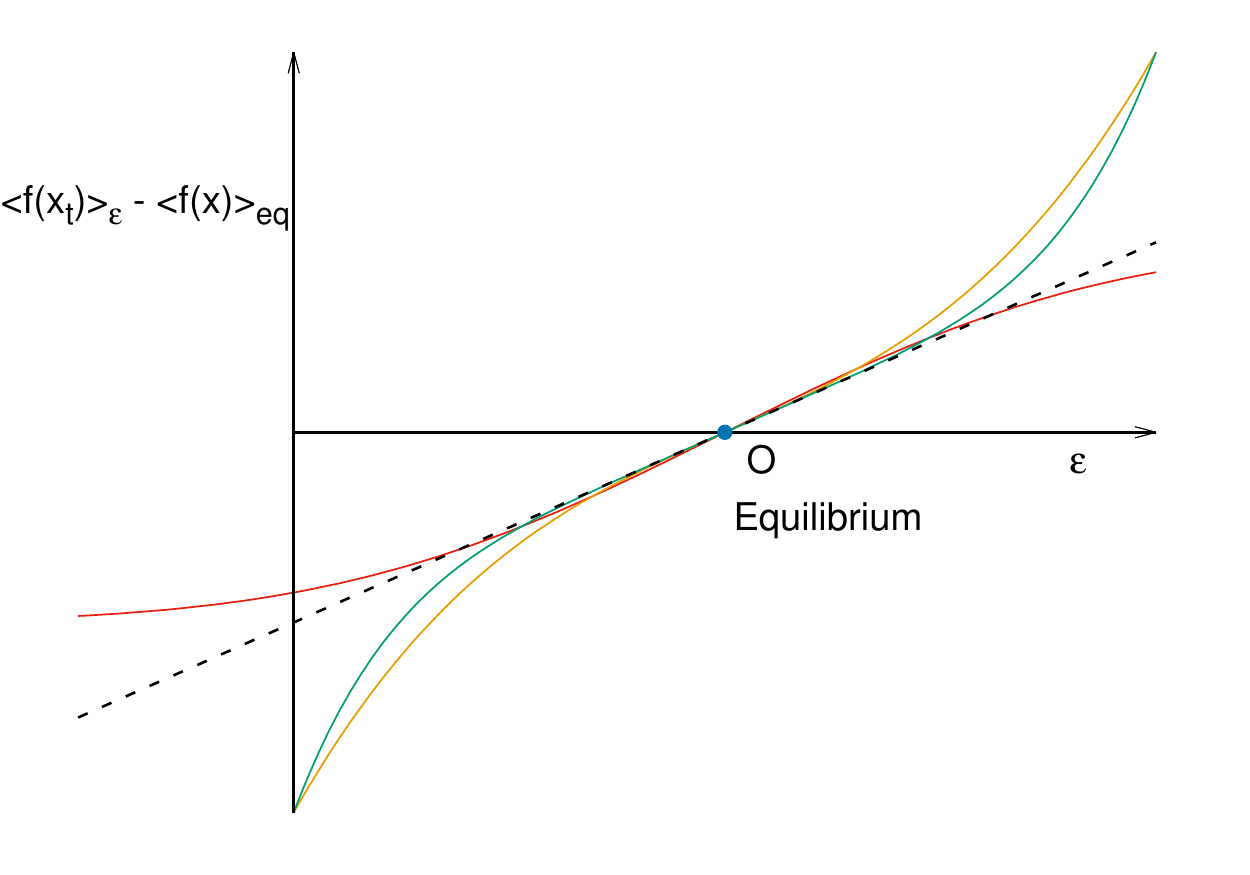}
	\caption{The scenario for nonlinear response around equilibrium.  The vertical axis shows {\it some} displacement as function of nonequilibrium (driving) paramter $\epsilon$.  The three functions correspond to different kinetics by which the same thermodynamic perturbation is realized.  In linear order, the responses coincide and deviations, much as life itself, start at second order around equilibrium.}
	\label{fig:sad}
\end{figure}

That situation changes in second order around equilibrium as seen from \eqref{kubo2}. We sketched the general scenario in Fig.~\ref{fig:sad}. The frenetic contribution enters and exit and entrance rates of the particles now matter.  The response has become sensitive to kinetic information beyond the change in (thermodynamic) chemical potential.  There are indeed different {\it kinetic} ways to increase the bath chemical potential and the difference will be picked up by the time-dependence of  $\langle N(t) \rangle - \langle N\rangle_\text{eq}$ in second order around equilibrium ($\delta^2$).  As first explored in \cite{pccp}, the total exchange activity (between the
system and the reservoir) enters, which is a time-symmetric traffic.

\subsubsection{Experimental challenges}\label{expf}

Second order response around equilibrium was explored first in \cite{urna16} for a colloidal particle in an anharmonic potential.  There, the technique to measure the trajectory of the particle is known as total internal
reflection microscopy.  The perturbation is an optical force on the particle.\\
In \cite{urna18} the problem of coarse-graining is investigated.  A trajectory-based response theory for a dense suspension is obviously challengiing.  As we saw before, also in Section \ref{open}, getting ``enough'' kinetic information in many-body systems is problematic to evaluate the frenetic contribution. Such coarse-graining aspects also can be studied in simulation and numerical studies.

\section{Quantum case}\label{qca}
The formalism of linear response theory as developed in the 1960's much followed that of perturbation theory in quantum mechanics. We repeat the main steps of that formalism, limiting ourselves to finite systems.  Mathematically rigorous generalizations to spatially-extended systems, to ground states in particular and to the description of linear response in the thermodynamic limit are obviously important, but today seem restricted to systems showing a mass gap uniformly in the volume; see e.g. \cite{woj}.\\
One starts with a Hamiltonian
\[
H(s) = H_0 + H_I(s),\qquad H_I(s) := -h_s\,B\]
where the operator $B$ stands for the perturbation, time-modulated with the small real $h_s=h_s^*$ (denoting complex conjugate).  The reference Hamiltonian is $H_0$. Associated to $H_0$ is the reference density matrix $\rho_0$, representing the initial mixed state before the perturbation sets in ($h_s=0$ for $s\leq 0$). From then on the dynamics is unitary as for a closed isolated system with evolution operator $U(s), s\geq 0$ satisfying
\[
i\hbar\, \frac{\id}{\id s}U(s) = H(s)U(s), \quad \text{ while } \; U_0(s) = e^{-i\frac{s}{\hbar} H_0}
\]
The initial density matrix $\rho_0$ is invariant for $U_0$: $U_0(s)\,\rho_0\,U_0^*(s) =\rho_0$.
A first order calculation gives
\[
U(t) = U_0(t) -\frac{i}{\hbar}\int_0^t\id s\, U_0(t-s)\,H_I(s)\, U_0(s) + O(h^2)
\]
or, in first order and with $B_0(u):= U_0^*(u)\,B\,U_0(u)$,
\[
U(t) = \left(1 + \frac{i}{\hbar}\int_0^t\id s \,h_s\,B_0(s-t)\right)\,U_0(t)
\]
That is all we need to calculate the density matrix $\rho(t), t>0$, to first order in $h_s$:
\begin{eqnarray}
\rho(t) &=& U(t)\,\rho_0\,U^*(t)\nonumber\\
&=& \rho_0 + \frac{i}{\hbar}\int_0^t\id\,h_s\,[B_0(s-t),\rho_0] + O(h^2)
\end{eqnarray}
We obtain the perturbed expectations from $\langle A(t)\rangle = \Tr[\rho(t)\,A]$ for observables $A$.  Writing
$A_0(t):= U_0^*(t)\,A\,U_0(t)$ we conclude that the response function is given by
\begin{eqnarray}\label{rq}
R_{AB}(t,s) &=& \frac{i}{\hbar}\Tr\big[\rho_0\,[A_0(t),B(s)]\big]\nonumber\\
&=& \frac{i}{\hbar}\Tr\big[[B,\rho_0]\,A_0(t)]\big]
\end{eqnarray}
for $t\geq s > 0$. That also works for ground states $\rho_0 = |0\rangle\langle0|$ (projector on the (nondegenerate) ground state of $H_0$):
\[
R_{AB}(t,s) = \frac{i}{\hbar}\,\langle 0|\,[A_0(t),B(s)]\,|0\rangle
\]
and obviously, by the stationarity of $\rho_0$, the response only depends on the time-difference $\tau=t-s>0$.\\

To reach the quantum fluctuation--dissipation theorem one must use that $\rho_0$ is the thermal equilibrium state for $H_0$. At this point one can use the Kubo-Martin-Schwinger condition for equilibrium densities $\rho_0=\rho_\text{eq} = \exp -\beta H_0/Z$, Tr$[\rho_\text{eq} A] = \langle A\rangle_\text{eq}$, which says
\[
\langle A(t-i\beta\hbar)\, B(t')\rangle_\text{eq} = \langle B(t') A(t)\rangle_\text{eq}
\]
That basically uses analyticity in a complex-time domain where $B_0(-i\hbar s)= 
e^{sH_0} \,B\,e^{-sH_0}$.  We thus have
\[
\int_0^{\beta\hbar} \id s\,\frac{\id B_0}{\id s}(-i s) = e^{\beta H_0}\,B\,e^{-\beta H_0} - B
\]
and \eqref{rq} becomes
\begin{equation}\label{qk}
R_{AB}(t,s) = \frac{i}{\hbar}\,\int_0^{\beta\hbar}\id\tau\,\left< \frac{\id B_0}{\id s}(-i\tau) \,A_0(t)\right >_\text{eq}
\end{equation}
which is the direct quantum analogue of the Kubo formula \eqref{kubs}.\\

Another approach takes the Fourier transform; see before in Example \ref{lav}.  One defines the equilibrium time-correlation
\begin{equation}\label{anti}
G_{AB}(t) := \frac 1{2}\langle A B_0(t) + B_0(t)A\rangle_\text{eq}
\end{equation}
where we can put that $\langle A\rangle = \langle B\rangle =0$ without loss of generality.  Assuming that the decay in time $t$ is sufficiently fast, we define the Fourier transform
\[
\tilde{G}_{AB}(\nu) = \int\id t \,G_{AB}(t)\,e^{i\nu t}
\]
where $\nu$ is the time-conjugate complex variable.  Since $G_{AB}(t)\in\bbR$, we have
\[
\tilde{G}^*_{AB}(\nu) = \tilde{G}_{AB}(-\nu), \quad \tilde{G}_{AB}(\nu) =\tilde{G}_{BA}(-\nu) 
\]
where the second equality follows from the cyclicity of the trace making $G_{AB}(t)=G_{BA}(-t)$.  In particular, $G_{AA}(t)$ is positive-definite, meaning that
\[
\sum_{i,j=1}^n c_i\,c_j^*\,G_{AA}(t_i-t_j) >0
\]
for all coefficients $c_i\in \bbC$.  That can be shown by using
\[
G_{AA}(t_i-t_j) = \frac 1{2}\Tr[\rho_0\,(A_0(t_i)A_0(t_j)+A_0(t_j)A_0(t_i))]\]
and it implies that $\tilde{G}_{AA}(\nu)\geq 0$ is real and positive.\\
A final calculation from \eqref{rq} leads to the fluctuation--dissipation theorem in the form
\begin{equation}\label{qfdr}
\frac 1{2i}\left(\tilde{R}_{BA}(\nu)-\tilde{R}^*_{AB}(\nu)\right) = \frac 1{\hbar}\tanh\left(\frac{\beta\hbar\nu}{2}\right)\,\tilde{G}_{AB}(\nu)
\end{equation}
That is the better known quantum version of the Kubo relation \eqref{kubs} (obtained from taking $\tanh(\beta\hbar\nu/2) \simeq \beta\hbar\nu/2$).\\  %We refer to \cite{sorkin} for an application to Brownian motion at absolute zero.\\
 When $A=B$, we have
\begin{equation}\label{aa}
\text{Im} \tilde{R}_{AA}(\nu) = \frac 1{\hbar}\tanh\left(\frac{\beta\hbar\nu}{2}\right)\,\tilde{G}_{AA}(\nu) >0, \quad \nu>0
\end{equation}
It is the imaginary part of the response function that relates to dissipation.  If indeed we consider  $E(t) =  \Tr(\rho(t)H(t))$ and we take $h_s = $Re$ (h_0e^{-i
\nu s}), A=B$, then
\[
E(2\pi/\nu) - E(0) = 
	\pi\,|h_0|^2\,\,\text{Im}\tilde{R}_{AA}(\nu)
	\]
	where the left-hand side is the change of energy over one period.  That dissipation is connected to fluctuations via the right-hand side of \eqref{aa}. In general one can find also the real part of the response by using the so called Kramers-Kronig relations,
	\begin{eqnarray*}
		\text{Re } \tilde{G}(\nu_0) = \frac 1{\pi} \int\id \nu \,\frac{\text{Im }\tilde{G}(\nu)}{\nu-\nu_0}\\
		\text{Im } \tilde{G}(\nu_0) = \frac 1{\pi} \int\id \nu \,\frac{\text{Re }\tilde{G}(\nu)}{\nu-\nu_0}
\end{eqnarray*}
where the integrals are for Principal Values.\\
Let us add that we can get rid of the ``Imaginary,'' say in \eqref{aa} by defining the odd response function
\[
R^o(\tau) = \text{sign}(\tau)\, R(|\tau|)
\] 
for which then $\tilde{R}^o_{AA}(\nu) = 2i\,\text{Im} \tilde{R}_{AA}(\nu)$, or 
\begin{equation}\label{snd}
\tilde{R}^o_{AA}(\nu) = \frac{2i}{\hbar}\tanh\left(\frac{\beta\hbar\nu}{2}\right)\,\tilde{G}_{AA}(\nu)
\end{equation}
and we can go back to the time-domain by taking convolutions.\\

The quantum version of the Sutherland--Einstein version is readily obtained from \eqref{aa}.  The mean square displacement is (using anti-commutators) 
\begin{eqnarray*}
\langle (X_t-X_0)^2\rangle_\text{eq} &=& \langle\{X_0,X_0\}\rangle_\text{eq} - \langle\{X_0,X_t\}\rangle_\text{eq}\\
&=& 2[G(0)-G(t)]
\end{eqnarray*}
where we inserted \eqref{anti} for $G(t) = \frac 1{2}\,\langle\{X_0,X_t\}\rangle_\text{eq}$.  Following \cite{sorkin}, with \eqref{aa} that implies that the diffusive behavior can be deduced from
\begin{eqnarray}\label{sor}
\langle (X_t-X_0)^2\rangle_\text{eq} &=&  \frac 1{\beta}\int_0^\infty\id \tau \, R(\tau)\left[2\coth(\pi\tau/\beta \hbar) -\coth(\pi(\tau+t)/\beta \hbar) - \coth(\pi(\tau-t)/\beta \hbar)\right]\nonumber\\
 &=& \frac{\hbar}{\pi}\int_0^\infty\id u \, R\left(\frac{\hbar\beta}{\pi}u\right)\left[2\coth(u) -\coth(u + \frac{\pi t}{\beta \hbar}) - \coth(u -\frac{\pi t}{\beta \hbar})\right]
\end{eqnarray}
For the time-dependent response function we use \eqref{rq},
\[
R(\tau) = \frac 1{i\hbar}\,\langle[X_0,X_\tau]\rangle,  \qquad \tau\geq 0
\]
(zero for $\tau<0$.)  In the long time, classical regime we must take $\beta\hbar\ll 1/\gamma$ with $1/\gamma$ the relaxation time for $R(\tau) \rightarrow \mu$ as $\tau\uparrow \infty$, with $\mu$ the mobility.  Then, \eqref{sor} yields $\mu = \beta\,{\cal D}$ as in the classical Sutherland--Einstein relation; see Example \ref{lav}.  In the long time quantum regime where we consider relaxation times shorter than $\beta\hbar$, other (intrinsic quantum) behavior may arise, as studied in \cite{sorkin}.\\

The reason for recalling the above is not only for completeness.  The calculations above give the standard approach to FDR of the first kind.  Note the difference in approach with all that went before. An extension to quantum nonequilibrium dynamics is therefore not obvious.  There are formal extensions as an open quantum system in various regimes evolves in time according
to a classical Markov dynamics.  Those regimes are characterized by terminology like fast decoherence, Coulomb blockade, fast repeated measurements, Zeno
regime,  etc. where, such as in Lindblad dynamics, the relaxation of the density matrix corresponds to the convergence of an associated classical Markov dynamics.
That is not what we are finally after of course; we want true quantum effects where nonlocality, nonMarkov-behavior and entanglement play a role.  It seems we are far from there (cf. the open problem in Section \ref{open}). The approach of the present paper so far fails as well, as we have no trajectory-based picture for open quantum systems. Note that the Feynman path-integrals do not refer to real trajectories.  Rather we believe that a useful extension of the Bohmian formulation of quantum mechanics to open systems is most promising to deal with the necessary ideas of (quantum) traffic or dynamical activity, even to start in the semiclassical realm \cite{ward}.  Ideas of unravelling of trajectories \cite{gis,gnei} or of classical representations of spin density evolutions \cite{sher,garh} go in that same direction.\\
On the other hand, much of today's research activity in quantum nonequilibrium physics uses either the Schwinger–Keldysh nonequilibrium Green function technique \cite{sw,ke} or the Feynman–Vernon influence functional approach \cite{fey}.  The calculations using time--dependent nonequilibrium Green functions are rather tedious however, and we fail to see a powerful conceptual framework. The Feynman–Vernon approach is useful for deriving (certain) master equations for the reduced density matrix, with most emphasis on bosonic (thermal) environments.   

\section{Conclusions and outlook}

The tools for observing and manipulating mesoscopic kinetics have been growing sensationally. We are therefore hopeful that a response theory based on checking trajectories is useful.  The relevant dynamical ensembles are governed by an action on path-space, where the weight of the various possible trajectories of the considered dynamical variables are decided by a competition between excesses in entropy flux and frenesy. Indeed, under local detailed balance the antisymmetric part in the action gives the total entropy flux (per $k_B$) into the environment, while the time-symmetric part becomes essential outside the close-to-equilibrium regime. That frenesy collects kinetic information such as in escape rates and dynamical activity.  New phenomena and modifications in Einstein and Sutherland-Einstein relations provide interesting new challenges for exploring the nonequilibrium world.
\vspace{1.7cm} 

\noindent {\bf Acknowledgment}:  Thanks to Tirthankar Banerjee for much appreciated help with the figures.  This research was supported in part by the International Centre for Theoretical Sciences during a visit for the program - Fluctuations in Nonequilibrium Systems: Theory and applications (Code:ICTS/Prog-fnsta2020/03) and by the Raman Research Institute, both in Bangalore.  I am grateful to Urna Basu and Anupam Kundu for the great hospitality.  

\end{document}